\documentclass[]{jetp} 
\twocolumn 
\usepackage{graphicx}
\usepackage{epsfig}
\usepackage{amsmath}
\usepackage{amssymb}

\newcommand{\sign}{\text{sgn}}
\newcommand{\St}{\text{St}}

\begin{document}
\English

\title{Kinetic processes in Fermi-Luttinger liquids} 

\author{Alex}{Levchenko} 

\affiliation{Department of Physics, University of Wisconsin--Madison, Madison, Wisconsin 53706, USA} 

\author{Tobias}{Micklitz}

\affiliation{Centro Brasileiro de Pesquisas F\'{i}sicas, Rua Xavier Sigaud 150, 22290-180, Rio de Janeiro, Brazil}

\abstract{In this work we discuss extensions of the pioneering analysis by Dzyaloshinski\v{i} and Larkin [Sov. Phys. JETP \textbf{38}, 202 (1974)] of correlation functions for one-dimensional Fermi systems, focusing on the effects of quasiparticle relaxation enabled by a nonlinear dispersion. Throughout the work we employ both, the weakly interacting Fermi gas picture and nonlinear Luttinger liquid theory to describe attenuation of excitations and explore the fermion-boson duality between both approaches. Special attention is devoted to the role of spin-exchange processes, effects of interaction screening, and integrability. Thermalization rates for electron- and hole-like quasiparticles, as well as the decay rate of collective plasmon excitations and the momentum space mobility of spin excitations are calculated for various temperature regimes. The phenomenon of spin-charge drag is considered and the corresponding momentum transfer rate is determined. 
We further discuss how momentum relaxation due to several competing mechanisms, viz. triple electron collisions, electron-phonon scattering, and long-range inhomogeneities affect transport properties, and highlight energy transfer facilitated by plasmons from the perspective of the inhomogeneous Luttinger liquid model. Finally, we derive the full matrix of thermoelectric coefficients at the quantum critical point of the first conductance plateau transition, and address magnetoconductance in ballistic semiconductor nanowires with strong Rashba spin-orbit coupling.\\

\textit{\underline{For the special issue of JETP devoted to the 90th birthday jubilee of Igor E. Dzyaloshinski\v{i}}.}}

\maketitle

\section{Introduction}

The concept of quasiparticles plays a central role in the condensed matter physics of strongly interacting many-body quantum systems \cite{AGD,Pains}. For instance, in the context of electrons in conductors, one typically views the quasiparticle states as those evolving from the free electron gas to a Fermi liquid when adiabatically turning on the interaction. In accordance with Landau theory \cite{Landau}, 
quasiparticles inherit some of the basic quantum numbers of bare electrons such as spin, charge, and momentum. Their respective dispersion relations as well as thermodynamical and kinetic properties may, however, differ significantly due to interaction-induced renormalizations. A crucial advantage of the quasiparticle picture is that residual interactions are assumed to be weak, and can be systematically and controllably addressed by means of perturbation theory. The central question related to the validity of the quasiparticle description concerns their lifetime $\tau_{\text{qp}}$. Indeed, in the process of scattering quasiparticles decay, and their mere notion remains meaningful only if attenuation is weak enough so that they can be considered as sufficiently long-lived collective excitations. In Fermi systems, the Pauli principle severely limits the phase space available for quasiparticle collisions. The low temperature decay rate can then be estimated from the Golden rule as  
\begin{equation}\label{tau-FL}
\tau^{-1}_{\text{qp}}(\varepsilon,T)\propto (\nu V_0)^2(\varepsilon^2+\pi^2T^2)/\varepsilon_F.
\end{equation}
In this expression, the excitation energy $\varepsilon=v_F(p-p_F)$ of a quasiparticle with momentum $p$ is counted from the Fermi energy $\varepsilon_F$,  $\nu$ is the density of states and $V_0$ is the characteristic strength of the short-range repulsive interaction.\footnote{Throughout the paper we use units with Planck and Boltzmann constants set to unity $\hbar=k_B=1$.} 
The dominant microscopic scattering channel leading to Eq. \eqref{tau-FL} involves quasiparticle decaying into three: another quasiparticle and a particle-hole pair. The amplitude for this process is proportional to $V_0$, hence, the dimensionless factor of $(\nu V_0)^2$ in the scattering probability entering Eq. \eqref{tau-FL}. The factor $\varepsilon^2$ is the phase space volume for scattering of a quasiparticle with energy $\varepsilon$ compatible with the conservation of total energy and momentum. At finite temperatures the smearing of states in the energy strip of order $\sim T$ per particle leads to the corresponding $T^2$ dependence of $\tau^{-1}_{\text{qp}}$. Higher-order processes involving $2n+1$ quasiparticles, namely $n>1$ electron-hole pairs, are usually neglected as their respective rate scales with higher powers of energy. In particular, at zero-temperature the rate for relaxation processes of a quasiparticle with energy $\varepsilon$ involving $n$ particle-hole pairs vanishes as $\tau^{-1}_{\text{qp}}\propto \varepsilon^{2n}$. One notable property of Eq. \eqref{tau-FL} is that it predicts the same relaxation time for particle-like and hole-like excitations. Another property is that the crossover from zero-temperature to finite-temperature relaxation is governed only by one scale, viz. when the excitation energy compares to the temperature itself $\varepsilon\sim T$.     
 
In addition to the quasiparticle relaxation, which is often viewed as an out-scattering rate from a particular quantum state, one may address a more general question of relaxation of a nonequilibrium quasiparticle distribution function. In kinetic theory such problem is 
typically analyzed in the framework of the linearized Boltzmann equation. The eigenvalues of the corresponding collision operator define relaxation times of different distribution function modes. In three dimensional Fermi liquids this problem is exactly solvable \cite{AK,BS} and one finds that all these rates are parametrically the same, scaling respectively as $\propto T^2$. In contrast, in two-dimensional Fermi liquids, kinematics of head-on collisions leads to a parametrically distinct relaxation of odd and even momentum harmonics of the distribution function, in particular $\tau^{-1}_{\text{even}}\propto T^2/\varepsilon_F$ while $\tau^{-1}_{\text{odd}}\propto T^4/\varepsilon^3_F$ \cite{Gurzhi,Levitov}.         
 
The role of dimensionality in quasiparticle relaxation becomes the most dramatic in one-dimension (1D). This special case of electron liquids can be experimentally realized in quantum wires of GaAs/AlGaAs heterostructure \cite{Auslaender} or carbon nanotubes \cite{Bockrath} when particle density is such that only the lowest sub-band of transversal modes is occupied. It further requires that
temperature is sufficiently low and sample purity is sufficiently high, so that thermally- and disorder-induced transitions to higher sub-bands are suppressed. In addition, edge modes formed at the boundaries of a 2D electron gas when placed in a strong magnetic field in the  integer or fractional quantum Hall regime \cite{Chang-PRL,Chang-RMP}, or edge states of 2D quantum spin Hall topological-insulators \cite{RRDu}, provide other distinct examples of, respectively, chiral and helical quantum 1D electron liquids. 

In principle, all these systems can be successfully described within the framework of Luttinger liquid theory \cite{Haldane,Giamarchi,Maslov}, which builds out of the Tomonaga-Luttinger (TL) model \cite{Tomonaga,Luttinger}. As is known form pioneering works \cite{Mattis-Lieb,Dzyaloshinskii-Larkin,Luther-Peschel}, in the asymptotic low-energy limit $\varepsilon/\varepsilon_F\ll1$, the key properties of the TL model are essenatially non-Fermi liquid like. A power-law anomaly manifests in the suppression of the single particle density of states  
\begin{equation}\label{DoS-TL}
\nu(\varepsilon)=\nu_0\left(\frac{|\varepsilon|}{v_Fp_\Lambda}\right)^{2g}\frac{\sin(\pi g)}{\pi g}\Gamma(1-2g),
\end{equation}
and collapse of the quasiparticle residue in the distribution function. At $T\to0$ that is
\begin{equation}\label{FD-TL}
n(\varepsilon)=\frac{\Gamma(\frac{1}{2}+g)}{2\sqrt{\pi}\Gamma(1+g)}
\left[1-\frac{\Gamma(\frac{1}{2}-g)}{\Gamma(\frac{1}{2}+g)}\left(\frac{|\varepsilon|}{v_Fp_\Lambda}\right)^{2g}\sign(\varepsilon)\right]
\end{equation}
where $\nu_0=1/(2\pi v_F)$, $\Gamma(z)$ is the Euler's gamma function,  and $p_\Lambda$ is the momentum cutoff of the model (parametrically $p_\Lambda\sim p_F$). In the simplest spinless version of the TL-model with short-ranged interaction, a single 
dimensionless coupling constant,   
\begin{equation}
g=\frac{1}{2}\left[\frac{1+\nu_0V_0}{\sqrt{1+2\nu_0V_0}}-1\right],
\end{equation}
can be related to the zero-momentum Fourier component of the bare interaction potential $V_0$. The limit of weak interaction corresponds to $g\ll1$ and Eqs. \eqref{DoS-TL}-\eqref{FD-TL} are valid for $g<1/2$.\footnote{In Ref. \cite{Dzyaloshinskii-Larkin} the limit of strong interactions, $g>1$/2, was also considered, including the scenario when coupling between fermions of the same chirality is different from coupling between fermions of different chirality. For additional details on the derivation of Eq. \eqref{FD-TL} see also Ref. \cite{Gutfreund-Schick}.} However, a direct attempt to apply Luttinger liquid theory to the question of quasiparticle lifetime meets formidable challenges. In a fermionic representation of the TL-model, elaborated explicitly by Dzyaloshinski\v{i} and Larkin \cite{Dzyaloshinskii-Larkin}, the electron self-energy vanishes on the mass shell in all orders of perturbation theory and, consequently, correlation functions assume power-law tails. These results, and the absence of relaxation, can be alternatively understood from the Mattis and Lieb \cite{Mattis-Lieb}, and Luther and Peschel \cite{Luther-Peschel} bosonization construction, which maps interacting 1D fermions to a collection of decoupled harmonic modes of charge-density and spin-density oscillations. Notably, in both approaches the exact solution relies heavily on the linearization of the fermionic dispersion relation, which is a cornerstone approximation.  

One is then left with the natural puzzle whether incorporating curvature of the dispersion relation into the TL-model would cure the issue and yield a finite lifetime of excitations, thus possibly restoring Fermi liquid like properties of the system. This line of reasoning can be also corroborated within the fermionic picture, noting that spectrum nonlinearity softens phase space restrictions for quasiparticle scattering, thus making their relaxation possible. 

Similarly, at the level of the bosonic description, nonlinear terms of the dispersion relation couple charge and spin modes thus enabling their decay. However, it was quickly recognized that curvature cannot be included perturbatively, and a naive expansion leads to spurious divergences. These and other related questions to 1D kinetics, including the connection between the two pictures of the fermion-boson duality, attracted significant recent interest. This has lead to the development of the nonlinear Luttinger liquid theory, also referred to as Fermi-Luttinger liquid (FLL) theory (see Refs. \cite{Deshpande-Nature,Imambekov-RMP} for comprehensive reviews and references herein). Specifically for the problem of quasiparticle relaxation in quantum wires, various scattering rates were calculated within different interaction models for both, spinless \cite{Samokhin,Rozhkov,Khodas-FL,Bagrets,Pereira-Affleck,Micklitz-Levchenko,Teber,Furusaki,Ristivojevic,Protopopov} and spin-$1/2$ fermions \cite{Balents-Egger,Karzig,Schmidt,Pereira-Sela,AL,Bard,Matveev}. In parts of the present work we review and extend these results. 

On the experimental forefront the hallmark signatures of Luttinger liquid behavior have been observed by means of various spectroscopic techniques. Namely, power-law anomalies in the density of states, tunneling conductance, and current-voltage characteristics \cite{Chang-PRL,Bockrath,Yao,Auslaender-ResTunLL}, spin-charge separation \cite{Yacoby,Jompol}, and charge fractionalization \cite{Steinberg,Kamata}. Besides GaAs quantum wires, carbon nanotubes, and edge modes, clear features of Luttinger liquid physics have been identified in many other systems such as bundles of NbSe$_3$ \cite{Slot} and MoSe \cite{Venkataraman} nanowires, polymer nanofibers \cite{Aleshin} and conjugated polymers at high carrier densities \cite{Yuen}, as well as atomically controlled chains of gold atoms on Ge surfaces \cite{Blumenstein}, just to name a few distinct examples. In the most recent report \cite{Barak}, relaxation processes in quantum wires were captured and bounds on the corresponding timescales were determined, thus providing measurements of quasiparticle properties beyond the paradigm of linear Luttinger liquid theory. In a parallel line of developments \cite{Eisenstein,Altimiras,Sueur,Khrapai}, cooling of nonequilibrium quasiparticles in quantum Hall edge fluids was measured and the corresponding lengths scales oforthermalization processes were quantified.      

The focus of this communication is on the description of elementary kinetic processes inducing relaxation in nonlinear Luttinger liquids and their emergent transport properties. Keeping forward scattering electron-electron interactions and accounting for nonlinear contributions to the electron dispersion, this theory is beyond the Dzyaloshinski\v{i}-Larkin theorem. The latter relax kinematic constraints and open phase space for multi-loop corrections to the electron self-energy, thereby providing a plethora of inelastic processes which affect equilibrium as well as nonequilibrium properties of the 1D quantum electron liquids. The rest of this work is structured as follows. Sec. \ref{sec:rates} focuses on the hierarchy of relaxation times in Fermi-Luttinger liquids. We present results beyond parametric estimates, including detailed computations of a number of experimentally relevant interaction models.\footnote{In part this material was summarized in Sec. IV of the extensive review in Ref. \cite{Imambekov-RMP}.} The complimentary kinetic equation approach, applied to the quasiparticle picture of a weakly interacting Fermi gas, and spin- and charge-excitations of a Luttinger liquid, are explored concurrently. We present numerical estimates for experimentally measured relaxation rates and provide detailed comparison to previous results. In Sec. \ref{sec:transport}, the temperature dependence of kinetic coefficients is calculated, accounting for extrinsic mechanisms of momentum relaxation due to phonons or long-range inhomogeneities. The contribution to heat transport mediated by plasmons in the inhomogeneous Luttinger liquid is elucidated. We devote parts of the discussion to the thermoelectric properties at the first plateau transition of the quantum conductance. Finally, we consider effects of strong spin-orbit coupling and magnetoconductance in ballistic semiconductor nanowires. In Sec. \ref{sec:summary}, we provide concluding remarks by sketching a broader picture, commenting on related topics as well as open questions relevant for chiral, helical, and spiral versions of 1D Fermi-Luttinger liquids. Several Appendices accompany our presentation in the main text, providing additional technical details of the presented analysis and formalism.


\section{Hierarchy of relaxation processes}\label{sec:rates}

The physics of quasiparticle relaxation in  1D quantum electron liquids is perhaps a surprisingly rich and complicated problem. In part this has to do with the fact that, in contrast to their higher dimensional counterparts, two-particle collisions, namely scattering processes with the emission of a single particle-hole excitation, do not result in finite relaxation rates. This statement pertains to generic dispersion relations, 
i.e. including curvature, and not only applies to models with linear dispersion. Indeed, kinematics of two-particle scattering in 1D is such that particles either keep or swap their momenta, but neither of these options causes relaxation. To allow for the redistribution of momenta and, at the same time, to comply with restrictions of conservations laws one necessarily needs to consider triple electron collisions, or alternatively, assume some extrinsic mechanisms. 

The analysis of 1D kinematics of multi-particle collisions resolving energy and momentum conservations reveals a variery of possible scattering events. They ultimately lead to a hierarchy of relaxation stages in the system and an emergent asymmetry between 
the relaxation of particle-like and hole-like excitations. All processes can be broken down into several distinct classes. First are the forward scattering processes with soft momentum transfer that involve either (i) all particles from the same branch, or (ii) particles from both branches such that all initial and final states are near the Fermi energy. Second are processes involving states deeper in the band. These latter are relevant for (iii) the drift of quasiholes and (iv) backscattering processes that change the number of right and left moving excitations before and after the collision. We will refer to thermalization when discussing relaxation processes that proceed without backscattering. These processes determine the lifetime of quasiparticles associated to the redistribution of excess energy, and affect thermal transport properties of the system. In contrast, the notion of equilibration will be used to refer to relaxation processes involving the backscattering of quasiparticles, which ultimately govern electrical transport properties.         

\subsection{Quasiparticle interaction model}

In the picture of a weakly nonideal Fermi gas, the probabilities of particle collisions can be calculated perturbatively in the interaction, 
employing the usual $\hat{T}$-matrix formalism \cite{Taylor}. 
Within the Golden Rule, the scattering rate, 
\begin{equation}\label{eq:W}
W=2\pi|A|^2\delta(E-E')\delta_{P,P'},
\end{equation}
is expressed in terms of the scattering amplitude $A$ of the corresponding quantum process. Here $E(E')$ and $P(P')$ label total energy and momentum of initial (final) states, and the delta-function $\delta(E-E')$ along with the Kronecker delta $\delta_{P,P'}$ enforce energy and momentum conservations. In the semiclassical limit, the three-particle amplitude $A$ was considered in Ref. \cite{Sirenko}. The generalization to the degenerate quantum limit was presented in the work of Ref. \cite{Lunde}, and exchange terms were carefully examined in Refs. \cite{Khodas-FL,Karzig,LRM}. The resulting amplitude takes the form 
\begin{equation}\label{eq:A}
A=\frac{1}{L^2}\sum_{\mathbb{P}\mathbb{P}'}\sign(\mathbb{P})\sign(\mathbb{P'})\frac{V_{p'_a-p_a}V_{p'_c-p_c}}{\varepsilon_{p_b}+\varepsilon_{p_c}-\varepsilon_{p_b+p_c-p'_c}}\Xi_{\sigma\sigma'}.
\end{equation}
Here $L$ is the system size and sums run over all possible permutations $\mathbb{P}$ of momenta $p_{i}$ with $i=1,2,3$ starting from the direct scattering process $(p_1,p_2,p_3)\to(p'_1,p'_2,p'_3)$ to all its exchange processes, with $\sign(\mathbb{P})$ accounting for the sign of the particular permutation (using the convention that $\sign(123) = +1$). Each permutation comes with a spin-dependent factor $\Xi_{\sigma\sigma'}=\delta_{\sigma_a\sigma'_a}\delta_{\sigma_b\sigma'_b}\delta_{\sigma_c\sigma'_c}$ reflecting particle exchange. In the spinless case, the amplitude has an  identical structure to Eq. \eqref{eq:A} with $\Xi_{\sigma\sigma'}\equiv 1$. The amplitude consists of 36 distinct terms that can be split into groups of six, each representing one direct and five exchange scattering processes, respectively. 
Technically speaking Eq. \eqref{eq:A} appears from the iteration of the $\hat{T}$-matrix, $\hat{T}=V+V\hat{G}_0\hat{T}$, to  second order in the bare two-particle interaction potential $V$. Here $\hat{G}_0$ is the resolvent operator (viz. the free particle Green's function) and $\varepsilon_p$ denotes the energy-momentum dispersion relation. 

For practical applications to quasiparticle scattering in quantum wires, it is sufficient to assume the simple dispersion of a parabolic band $\varepsilon_p=p^2/2m^*$ with effective mass $m^*$, and use a Coulomb interaction potential. Effects of screening due to nearby gates can be modeled by a conducting plate placed at a distance $d$ away from the wire. In this case the interaction potential is of the form 
\begin{equation}
V(x)=\frac{e^2}{\varkappa}\left[\frac{1}{|x|}-\frac{1}{\sqrt{x^2+4d^2}}\right],
\end{equation}
where $\varkappa$ is the dielectric constant of the host material. The diverging short-range behavior of this potential needs to be regularized in order to evaluate the small-momentum Fourier components $V_p$ entering the amplitude in Eq. \eqref{eq:A}. To this end,  
we introduce the small width $w$ of the quantum wire, $w\ll d$, and replace $1/|x|\to1/\sqrt{x^2+4w^2}$. Upon 1D Fourier transform we then find 
\begin{equation}\label{V}
V_p=(2e^2/\varkappa)\left[K_0(2w|p|)-K_0(2d|p|)\right],
\end{equation}  
where $K_0(z)$ is the modified Bessel function of the second kind. Using the asymptotic expression of the Bessel function at $z\ll1$, $K_0(z)\approx\ln(2/ze^{\gamma_E})+\frac{z^2}{4}\ln(2/ze^{\gamma_E-1})$, with $\gamma_E$ the Euler constant, one then finds the simplified form 
of the interaction potential 
\begin{equation}\label{eq:V-SC}
V_p\approx(2e^2/\varkappa)\left[\ln(d/w) - (pd)^2\ln(e^{1-\gamma_E}/|p|d)\right],
\end{equation} 
applicable to the screened limit of Coulomb interaction and valid for $p\ll 1/d$.
In the opposite regime, $d^{-1}\ll p\ll w^{-1}$, the second term in Eq. \eqref{V} can be neglected since $K_0(z)\propto e^{-z}/\sqrt{z}$ at $z\gg1$. One then arrives at the simplified form of the unscreened potential
\begin{equation}\label{eq:V-C}
V_p\approx(2e^2/\varkappa)\left[\ln(e^{-\gamma_E}/|p|w) + (pw)^2\ln(e^{1-\gamma_E}/|p|w)\right].
\end{equation}

A few comments are in order in relation to the interaction model presented in this section. (i) It should be noted that retaining numerical pre-factors of the order of unity under the logarithm in above expressions for $V_p$ would exceed the accuracy of further calculations, so they will be dropped and simply set to unity. (ii) However, retaining the sub-leading corrections containing $p^2$ in the main log-series expansion of both Eqs. \eqref{eq:V-SC} and \eqref{eq:V-C} is actually crucial. Indeed, in the spinless case, the model with contact interaction as well as the Calogero-Sutherland model, are known to be completely integrable \cite{Sutherland}. This implies that all irreducible multi-particle scattering amplitudes must vanish identically for a constant $V_p$ and $V_p\propto |p|$. Furthermore, the extended model of short-ranged interaction, $V_p\propto p^2$, corresponding to the real space potential $V(x)\propto \delta''(x)$, is also integrable. This is known as Cheon-Shigehara model \cite{Cheon-Shigehara}. It is only due to the additional logarithm $\propto p^2\ln|p|$ in Eq. \eqref{eq:V-SC}, that there is partial non-cancellation between different terms in Eq. \eqref{eq:A} and the amplitude remains finite. (iii) In the model of long-ranged Coulomb interaction the situation is more subtle. A priori this model is not known to be integrable. Nevertheless, the amplitude in Eq. \eqref{eq:A} vanishes for pure logarithmic interaction $V_p\propto \ln|p|$, so that retaining an additional $p^2\ln|p|$ term in Eq. \eqref{eq:V-C} is important to get a finite result.   

The triple electron scattering rate from Eq. \eqref{eq:W} generates the collision integral (Stosszahlansatz) of the corresponding Boltzmann equation 
\begin{align}\label{eq:St-eee}
\!\!\St\{n\}&=\!\!\!\sum_{\{p\},\{\sigma\}}\!\!\!
W [n_{p'_1}(1-n_{p_1})n_{p'_2}(1-n_{p_2})n_{p'_3}(1-n_{p_3})\nonumber \\ 
&-n_{p_1}(1-n_{p'_1})n_{p_2}(1-n_{p'_2})n_{p_3}(1-n_{p'_3}].
\end{align}
Here each pair of Fermi functions, $n_p(1-n_{p'})$, captures statistical occupation probabilities, whereas the two terms of the collision integral correspond to incoming and outgoing processes. At thermal equilibrium these terms nullify each other by virtue of the detailed balance condition. At weak disequilibrium, one can linearize $n_p=f_p+\delta n_p$ in the external perturbation $\delta n_p$ around the equilibrium Fermi-Dirac distribution function $f_p$. The collision term can then be considered as a linear integral operator, acting on $\delta n_p=f_p(1-f_p)\psi$, and one can formulate the eigenvalue problem for this operator, $\St\{\psi_n\}=\omega_n\psi_n$. The spectrum of eigenvalues $\omega_n$ may be discrete or continuous, and captures all the information about the decay of different distribution function modes. As solving this problem exactly for triple collisions presents a daunting task \cite{Micklitz-Levchenko,Matveev}, we here follow a simpler more pragmatic approach. Setting, for instance, $\delta n_{p_1}=\delta_{p_1,p_F+\varepsilon/v_F}$ describes an injected quasiparticle with excess energy $\varepsilon$. Neglecting then secondary collisions, the Boltzmann equation reduces to the simple relaxation time approximation, $(\partial_t+\tau^{-1}_{\text{qp}})\delta n_p=0$, with solution $\delta n_p\propto\exp(-t/\tau_{\text{qp}})$. 
It is natural to identify the corresponding timescale for decay with the quasiparticle life-time   
\begin{equation}
\tau^{-1}_{\text{qp}}=-\partial\text{St}\{n\}/\partial n_p,
\end{equation}
which follows from Eq. \eqref{eq:St-eee} by only retaining the out-scattering contribution. Alternatively, one may project the collision operator \eqref{eq:St-eee} onto either momentum or energy modes and thus infer the relaxation time of interest. This approach is parametrically correct, however, may miss numerical factors of  order unity when compared to the exact solution of the eigenvalue problem. 
We will employ both approaches in the forthcoming sections. 

\subsection{Quasiparticle decay rates}\label{sec:qp-decay}

Owing to one-dimensionality of the problem, it is convenient to think of particles of different chirality, namely right-movers (R) and left-movers (L). It can be readily checked that strictly at zero-temperature quasiparticle relaxation is only possible if collisions involve both, right- and left-moving particles since otherwise conservation laws cannot be satisfied. For this reason, consider first a process of relaxation that involves two right-moving particles, with initial momenta $p_1,p_2$, and a left-moving particle labeled by momentum $p_3$. The outgoing momenta after the collision, $p'_i=p_i+q_i$, will be labeled by momenta transfer $q_i$ for each of the particle $i=1,2,3$. In these notations, the momentum conservation becomes $q_1+q_2+q_3=0$, and the energy conservation, for a simple parabolic band, can be cast in the form $2(p_1q_1+p_2q_2+p_3q_3)+q^2_1+q^2_2+q^2_3=0$. These conditions set the phase-space constraints for collisions. 

For an initial state with $p_1=p_F+\varepsilon/v_F$, the quasiparticle life-time corresponding to an RRL-process is then 
\begin{equation}
\tau^{-1}_{\text{qp}}=\sum_{\stackrel{p_2p_3}{p'_1p'_2p'_3}} W(1-f_{p'_1})f_{p_2}(1-f_{p'_2})f_{p_3}(1-f_{p'_3}),
\end{equation}
where we first focused on a simpler case of spinless electrons. At this point it is convenient to shift momenta of left- and right-movers from the respective Fermi points, $p_{1,2}=p_F+k_{1,2}$ and $p_3=-p_F+k_3$. In addition, it is sufficient to linearize the spectrum in the distribution functions, approximating $f_{\pm p_F+k}\to f_{\pm k}=(e^{\pm v_Fk/T}+1)^{-1}$, but not in the scattering probability $W$. 
Indeed, an analysis of the kinematic constraints suggests that $q_1\approx-q_2$ and $q_3\approx(q_1/p_F)(k_1-k_2+q_1)$, implying that $|q_3|\ll |q_{1,2}|$. In other words, relaxation occurs in incremental steps of momentum transfer $q_3\sim \varepsilon^2/v^2_Fp_F$ from right-movers to left-movers. With these observations at hand, we next need the corresponding three-particle scattering amplitude. For the case of long-ranged Coulomb interaction Eq. \eqref{eq:V-C}, one finds from Eq. \eqref{eq:A} after a laborious expansion  
\begin{equation}\label{eq:A-C}
A\approx\frac{2(p_Fw)^2}{L^2\varepsilon_F}\left(\frac{2e^2}{\varkappa}\right)^2
\left[1-\frac{3}{4}\ln\left(\frac{1}{p_Fw}\right)\right]\ln\left(\frac{q^2_1}{p_F|q_3|}\right).
\end{equation} 
This result is obtained to the leading logarithmic accuracy using two small parameters $|q_1|/p_F\sim|q_3|/|q_1|\ll1$ in the expansion. 
With the same level of accuracy the momentum and energy conservations in Eq. \eqref{eq:W} can be simplified to
\begin{equation}
\delta_{P,P'}\delta(E-E')\approx\frac{1}{v_F}\delta(q_3-[q_1(k_1-k_2)+q^2_1]/p_F)\delta{q_1,-q_2}.
\end{equation}
These approximations enable one to complete all five momentum integrations. Two integrations are removed by delta functions fixing values of $q_2$ and $q_3$ in terms of $k_{1,2}$ and $q_1$. Furthermore, in the zero temperature limit, $T\to0$, Fermi occupations become step-functions, $f_k\to\theta(-k)$. The integral over $k_3$ then becomes elementary, contributing by a pure phase space factor
$\sum_{k_3}f_{-k_3}(1-f_{-k_2-q_3})=(L/2\pi)|q_3|\theta(-q_3)$.  The product of Fermi factors, $f_{k_2}(1-f_{k_2-q_1})$, simply limits the domain of $k_2$ to the range $k_2\in [-|q_1|,0]$, while the remaining $1-f_{k_1+q_1}$ dictates that $q_1<k_1$, and we
recall that in this setting $k_1=\varepsilon/v_F$. Putting everything together the RRL-process gives the life-time
\begin{equation}\label{eq:tau-spinless-C}
\tau^{-1}_{\text{qp}}=c_1\varepsilon_Fg^4\lambda^2_{1}(p_Fw)(\varepsilon/\varepsilon_F)^4,
\end{equation}
where $g=e^2/\varkappa v_F$ is the dimensionless parameter of interaction strength in the model, and we introduced notation for $\lambda_1(z)=z^2\ln(1/z)$. The numerical coefficient $c_1=(15-\pi^2)/32\pi^3$ is obtained with help of the integral
\begin{equation}
\iint^{1}_{0}x^2q(x,y)\ln^2\left[x/q(x,y)\right]dxdy=\frac{15-\pi^2}{72},
\end{equation}
where $q(x,y)=1-x(1-y)$. We here notice that the numerical factor $c_1$ in Eq. \eqref{eq:tau-spinless-C} differs 
from that calculated in Refs. \cite{Khodas-FL,Pereira-Affleck} as different properties of the 
interaction potential were assumed.\footnote{In Appendix we sketch derivation of Eq. \eqref{eq:tau-spinless-C} from the bosonization framework of a mobile impurity scattering in Luttinger liquids.}

We see that finite decay rate emerges in forth order of the interaction strength. We also notice that the attenuation is inversely proportional to the cube of mass, $\tau^{-1}_{\text{qp}}\propto (m^*)^{-3}$, and vanishes as the limit $m^*\to\infty$ is taken at fixed band velocity.  
This limit corresponds to the situation considered by Dzyaloshinski\v{i} and Larkin. The energy scaling of the decay rate, $\propto \varepsilon^4$, is consistent with expectations based on the Fermi liquid picture for a process involving two particle-hole pairs. However, this result is not universal. This becomes evident from repeating the above calculation for the model of screened short-range interaction, i.e. using the potential given by Eq. \eqref{eq:V-SC}. Expanding the amplitude in Eq. \eqref{eq:A} under the same conditions as above, one then finds instead of Eq. \eqref{eq:A-C} the amplitude
\begin{align}\label{eq:A-SC}
&A\approx-\frac{5(p_Fd)^4}{3L^2\varepsilon_F}\left(\frac{2e^2}{\varkappa}\right)^2\ln\left(\frac{1}{p_Fd}\right)\nonumber \\ 
&\times\left[\frac{q^2_1}{4p^2_F}\left[1+6\ln\left(\frac{|q_1|}{p_F}\right)\right]-\frac{q^2_3}{q^2_1}\left[1+6\ln\left(\frac{|q_3|}{|q_1|}\right)\right]\right].
\end{align}
The crucial difference here compared to Eq. \eqref{eq:A-C} is the appearance of the additional small parameter $|q_1|/p_F\sim|q_3|/|q_1|\sim\varepsilon/\varepsilon_F\ll1$, which can be related to the fact that this particular model  is nearly integrable.
A close inspection of the amplitude in Eq. \eqref{eq:A} reveals that each term individually diverges as $1/q$ at small characteristic momentum transfer. However, all exchange terms combined together remove the singularity and partially cancel out all the way to $\sim q^2\ln q$ order. The rest of the calculation carries through in exactly the same way as in the previous example, and one finds the decay rate
\begin{equation}\label{eq:tau-spinless-SC}
\tau^{-1}_{\text{qp}}=c_2\varepsilon_Fg^4\lambda^2_2(p_Fd)(\varepsilon/\varepsilon_F)^8\ln^2(\varepsilon_F/\varepsilon)
\end{equation}
with $c_2=2445/3584\pi^3$ and $\lambda_2(z)=z^4\ln(1/z)$. The four extra powers in the energy dependence, can be traced back to the different asymptotic form of the amplitude in Eq. \eqref{eq:A-SC}. This demonstrates high sensitivity of decay rates in 1D systems to details of the interparticle interaction. The result captured by Eq. \eqref{eq:tau-spinless-SC} is of course perturbative. For generic nonintegrable models with short-ranged interaction, it can be generalized to arbitrary interaction strength. It can further be shown that $\tau^{-1}_{\text{qp}}\propto \varepsilon^8$ remains valid, and the pre-factor can be expressed in terms of the exact spectrum, see Ref. \cite{Furusaki} for details.   

As should be anticipated from the discussion above, electron spin plays a crucial role in the transition matrix element for the three-particle process, and thus significantly affect the quasiparticle decay rate. Indeed, in the spinless case antisymmetry of the electron wave function dictates that its orbital component should be odd and therefore relevant exchange amplitudes are suppressed by Pauli exclusion. Mathematically, one sees this in a cancellation of various terms that lead to Eq. \eqref{eq:A-SC}. In contrast, for spinful  
electrons, singular parts of the amplitude do not cancel. They are dominated by  $2p_F$ exchange-processes 
between branches, in which left-movers are scattered into right-movers \cite{Karzig}. Even though the strength of $2p_F$ exchange interaction is weaker than that at small momentum scattering, $V_{2p_F}\ll V_0$, for Coulomb interaction the relative reduction is only logarithmic. The gain in the amplitude, on the other hand, is much more substantial and controlled by the large factor $\sim\varepsilon_F/v_Fq\gg1$. This statement can be verified explicitly from Eq. \eqref{eq:A} where after spin summation one finds for the square of the amplitude of the RRL process 
\begin{equation}\label{eq:A-spinful}
\sum_{\stackrel{\sigma_2\sigma_3}{\sigma'_1\sigma'_2\sigma'_3}}|A|^2=\frac{3V^2_{2p_F}(V_0-V_{2p_F})^2}{32L^4\varepsilon^2_F}\left[\frac{q^2_1}{q^2_3}+\frac{4p^2_F}{q^2_1}\right].
\end{equation}  
To obtain this result we approximated $V_{p_1-p_2\pm q_i}\approx V_0$ and $V_{p_{1,2}-p_3\pm q_i}\approx V_{2p_F}$ in all the relevant terms since $p_{1,2}-p_3\approx 2p_F$ and $q_i\ll |p_i|$. Again, by repeating momentum integrations, the decay rate is found to be of the form  
\begin{equation}\label{eq:tau-spinful}
\tau^{-1}_{\text{qp}}=c_3\varepsilon_Fg^4\lambda^2_3(p_Fw)(\varepsilon/\varepsilon_F)^2\ln^2(\varepsilon_F/\varepsilon),
\end{equation}
with $c_3=45/32\pi^3$ and $\lambda_3(z)=\ln(1/z)$. To be consistent with the approximations that lead to Eq. \eqref{eq:A-spinful}, the difference $V_0-V_{2p_F}$ should be understood as a weak logarithmic factor $\simeq (2e^2/\varkappa)\ln(\varepsilon_F/\varepsilon)$ for the Coulomb interaction potential. This was incorporated into Eq. \eqref{eq:tau-spinful}. The singularity of the amplitude was compensated by
phase space factors, and perhaps surprisingly this restores essentially the Fermi liquid form of the decay rate at $T=0$. We note that up to 
model dependent pre-factors, the quadratic dependence of the relaxation rate on energy of spin-$1/2$ particles given by Eq. \eqref{eq:tau-spinful} is consistent with predictions of previous studies \cite{Karzig,Bard}.   

We proceed with a discussion of the effects of thermal broadening on relaxation processes. In the Fermi liquid picture one 
expects a simple crossover at excitation energies of the order of temperature $\varepsilon\sim T$. For 1D liquids this is not the case, as 
even at $T<\varepsilon$ there are intermediate regimes and relaxation shows nontrivial temperature dependence. 
Indeed, at finite temperatures each collision results in a typical momentum transfer $q_i\sim T/v_F$ allowed by thermal smearing of states near the Fermi energy. As RRL relaxation is controlled by momentum transfer between the branches, one needs to compare 
phase spaces available to left movers. Since at zero temperature $q_3\sim \varepsilon^2/v^2_Fp_F$, one deduces
from comparison to $q_3\sim T/v_F$ the crossover scale $\varepsilon_T\sim\sqrt{\varepsilon_FT}$. 
Technically, this argument also becomes evident observing that $\sum_{k_3}f_{-k_3}(1-f_{-k_3-q_3})=(L/2\pi)q_3(e^{v_Fq_3/T}-1)^{-1}$, and reducing to $LT/2\pi v_F$ as $q_3\to0$. These considerations suggest that Eqs. \eqref{eq:tau-spinless-C}, \eqref{eq:tau-spinless-SC}, and \eqref{eq:tau-spinful} are valid for $T\ll \varepsilon^2/\varepsilon_F$. Above this threshold one finds 
\begin{equation}\label{eq:tau-spinless-C-T}
\tau^{-1}_{\text{qp}}=c_4\varepsilon_Fg^4\lambda^2_1(p_Fw)(\varepsilon/\varepsilon_F)^2(T/\varepsilon_F),
\end{equation} 
instead of Eq. \eqref{eq:tau-spinless-C} for the spinless Coulomb case. 
Similarly,
\begin{equation}\label{eq:tau-spinless-SC-T}
\tau^{-1}_{\text{qp}}=c_5\varepsilon_Fg^4\lambda^2_2(p_Fd)(\varepsilon/\varepsilon_F)^6(T/\varepsilon_F)\ln^2(\varepsilon_F/\varepsilon),
\end{equation}
instead of Eq. \eqref{eq:tau-spinless-SC} for the spinless screened case, and finally 
\begin{equation}\label{eq:tau-spinful-T}
\tau^{-1}_{\text{qp}}=c_6\varepsilon_Fg^4\lambda^2_3(p_Fw)(T/\varepsilon_F)\ln^2(\varepsilon_F/\varepsilon)
\end{equation}
instead of Eq. \eqref{eq:tau-spinful} for the spin-$1/2$ Coulomb case. 
The set of coefficients $c_{4,5,6}$ can be determined from numerical integrations, however, 
their specific values are  of no particular significance here.     

At elevated temperatures the above mechanism for relaxation competes with another process involving only particles of the same chirality. As indicated earlier, this RRR- (or equivalently LLL-) process is kinematically possible only at finite energies. It follows from the
same amplitude Eq. \eqref{eq:A}, but admits different conditions on the involved momenta. In this process, a high-energy particle with excess energy $\varepsilon$ can relax on two other comoving particles, which during the collision
are scattered in opposite directions in energy. Namely,  one is drifting slightly upwards in energy, whereas the other float downwards, closer to the Fermi point. A detailed calculation in the spinless Coulomb model shows that the corresponding relaxation rate is given by 
\begin{equation}
\tau^{-1}_{\text{qp}}=c_7g^4(p_Fw)^4(T^3/\varepsilon\varepsilon_F)\ln^2(\varepsilon_w/\varepsilon),
\end{equation} 
where $\varepsilon_w=v_F/w$. This rate exceeds that given in Eq. \eqref{eq:tau-spinless-C-T}, provided that temperature is higher than $\sim \varepsilon\sqrt{\varepsilon/\varepsilon_F}$. In the case of screened Coulomb interaction, the same mechanism is more strongly suppressed 
 \begin{equation}
\tau^{-1}_{\text{qp}}=c_8g^4(p_Fd)^8(T^7/\varepsilon\varepsilon^5_F)\ln^2(\varepsilon_d/\varepsilon)\ln^2(\varepsilon/T),
\end{equation} 
 where $\varepsilon_d=v_F/d$. In fact, $\propto T^7$ is a generic property for any non-integrable finite-range interaction model with a sufficient degree of analyticity at small momenta \cite{Ristivojevic,Protopopov}. Lastly, in the case of spin-$1/2$ chiral electrons one estimates the decay rate to be of the form 
 \begin{equation}
  \tau^{-1}_{\text{qp}}=c_9g^4(T\varepsilon^6_T/\varepsilon^2\varepsilon^4_d)\ln^4(d/w). 
\end{equation}

\begin{table}
\begin{tabular}{ |c|c|c|c| } 
 \hline
 $\tau^{-1}_{\text{qp}}$ & $T<T_1$ & $T_1<T<T_2$ & $T_2<T<\varepsilon$\\ 
 \hline\hline
 Coulomb & $\varepsilon^4/\varepsilon^3_F$ & $T\varepsilon^2/\varepsilon^2_F$ & $T^3/\varepsilon\varepsilon_F$\\ \hline
 Screened & $\varepsilon^8/\varepsilon^7_F$ &$T\varepsilon^6/\varepsilon^6_F$ & $T^7/\varepsilon\varepsilon^5_F$ \\ \hline 
 Spin-$1/2$ & $\varepsilon^2/\varepsilon_F$ & $T$ & $T\varepsilon^6_T/\varepsilon^2\varepsilon^4_d$ \\ 
 \hline
\end{tabular}
\caption{Energy and temperature dependencies of
quasiparticle relaxation rates (only the leading parametric behavior is indicated and logarithmic terms are omitted for brevity). 
First two rows summarize results for spinless electrons interacting via Coulomb and screened short-range interaction models, respectively, and the last row gives the result for the spin-$1/2$ fermions. The first two columns describe processes involving particles of both chiralities (e.g. the RRL process), and the last column describes
the relaxation of comoving particles with only same chirality (e.g. the RRR process). 
In all cases $T_1\sim \varepsilon^2/\varepsilon_F$, while $T_2\sim\varepsilon\sqrt{\varepsilon/\varepsilon_F}$ in the Coulomb model, $T_2\sim\varepsilon\sqrt[6]{\varepsilon/\varepsilon_F}$ in the screened model, 
and $T_2\sim\varepsilon(\varepsilon_d/\varepsilon_F)\sqrt[3]{\varepsilon_d/\varepsilon}$ in the spinful model.}\label{tbl}
 \end{table}

In addition to relaxation of particles with the same chirality, thermal broadening allows for the relaxation of hot quasiholes, a process  kinematically forbidden at zero temperature. The derivation of the corresponding decay rate $\tau^{-1}_{\text{qh}}$ proceeds in close analogy to that for the RRL process. Crucial modifications are (i) the sign of $q_3$, (ii) a smaller phase space volume, now suppressed by an additional factor $\sim T/(\varepsilon^2/\varepsilon_F)$, and (iii) that it takes $\sim (\varepsilon/\varepsilon_T)^2$ steps to relax the excess energy. As a result, the quasihole relaxation rate e.g. for the spin-$1/2$ model,
\begin{equation}
\tau^{-1}_{\text{qh}}=c_{10}\varepsilon_Fg^4\lambda^2_3(p_Fw)(T/\varepsilon)^2\ln^2(\varepsilon_F/\varepsilon),
\end{equation}
is by a factor $(\varepsilon_T/\varepsilon)^4$ smaller than $\tau^{-1}_{\text{qh}}$ defined in Eq. \eqref{eq:tau-spinful} when taken at the same energy. This pronounced asymmetry in the relaxation rates of electron-like and hole-like excitations is a direct consequence of the 1D kinematics of three-particle scattering with nonlinear spectrum. This feature marks a sharp distinction between the quantum 1D Fermi-Luttinger liquids and higher dimensional Fermi liquid counterparts. 

We summarize in Table~\ref{tbl}  the discussed quasiparticle relaxation rates 
in the different regimes and models. 
 
\subsection{Distribution imbalance rates}

Another common technique in kinetic theory applied to the determination of relaxation rates is to project the collision integral onto specific modes of interest, to infer their corresponding decay times. For instance, in the context of the present problem, one can look at the thermal imbalance relaxation between left- and right-movers. This amounts to projecting the collision term onto the energy mode of the distribution function $n_p$, which is even in momentum. 

To see the practical implementation of this method, consider a situation in which right-movers are hotter than left-movers.  
The goal is then to derive an  equation which describes the relaxation of the difference in temperatures $\Delta T=T^R-T^L$ of left- and right-moving electrons. It should be noted that the physical setting with imbalanced temperature is justified in 1D: while three-particle collisions generate both right- and left-moving particle-hole pairs the intrabranch relaxation induced by these processes is faster, 
while interbranch is slow. 

We start from the Boltzmann equation, multiply both sides by $\varepsilon_{p_1}-\varepsilon_F$, and sum over $p_1>0$
\begin{equation}\label{eq:BKE-T-relax}
\sum_{p_1>0}(\varepsilon_{p_1}-\varepsilon_F)\partial_tn_{p_1}=\sum_{p_1>0}(\varepsilon_{p_1}-\varepsilon_F)\St\{n\},
\end{equation}
where, as above, momentum $p_1$ is that of a right-moving particle.  We then assume $n_{p_1}$ to be of Fermi-Dirac form with nonequilibrium temperature $T^R=T+\Delta T$ of right-moving excitations, and linearize above equation in the left-hand-side with respect to $\Delta T$,
\begin{equation}
\partial_t n_{p_1}=\partial_Tn_{p_1}\partial_t\Delta T=\frac{(\varepsilon_{p_1}-\varepsilon_F)\partial_t\Delta T}{4T^2\cosh^2(\frac{\varepsilon_{p_1}-\varepsilon_F}{2T})}. 
\end{equation}
When computing the integral over $p_1$ it is convenient to shift momentum to the respective Fermi point, $p_1=p_F+k_1$. Linearizing further the dispersion relation in $k_1$, $\varepsilon_{p_1}-\varepsilon_F\approx v_Fk_1$, one may use that  
 the integral is peaked at $p_F$ and rapidly converging. Noting that $\int^{+\infty}_{-\infty}z^2dz/\cosh^2(z)=\pi^2/6$, one readily finds 
\begin{equation}
\sum_{p_1>0}(\varepsilon_{p_1}-\varepsilon_F)\partial_tn_{p_1}=\frac{\pi LT}{6v_F}\partial_t\Delta T.
\end{equation}

The next step is to also linearize the right-hand-side of Eq. \eqref{eq:BKE-T-relax} in $\Delta T$. To accomplish this task we parametrize $n_p=f_p+f_p(1-f_p)\psi_p$, which allows to conveniently take advantage of the detailed balance condition in the collision integral 
$\St\{n\}$. For the thermal imbalance $\psi_p=(\varepsilon_p-\varepsilon_F)\Delta T/T^2$, and one finds upon expansion in $\Delta T$  
\begin{equation}
\sum_{p_1>0}(\varepsilon_{p_1}-\varepsilon_F)\St\{n\}=-\frac{\Delta T}{T^2}\sum_{\{k,q,\sigma\}}
(v_Fk_1)(v_Fq_3)\mathcal{W}.
\end{equation} 
Here 
\begin{equation}
\mathcal{W}=Wf_{k_1}(1-f_{k_1+q_1})f_{k_2}(1-f_{k_2+q_2})f_{-k_3}(1-f_{-k_3-q_3}),
\end{equation}
and at intermediate steps we made use of the energy conservation implicit in $W$, and approximated $\varepsilon_{p_1}-\varepsilon_F\approx v_Fk_1$ and $\varepsilon_{p'_3}-\varepsilon_{p_3}\approx-v_Fq_3$. 
It is now evident that Eq. \eqref{eq:BKE-T-relax} can be cast in form of the usual relaxation time approximation,
\begin{equation}
\partial_t\Delta T=-\Delta T/\tau_{\text{th}},
\end{equation} 
where we introduced the corresponding thermalization time. For the kinematics of 
the RRL process, the latter evaluates to 
\begin{equation}
\tau^{-1}_{\text{th}}=c_{11}\varepsilon_Fg^4\lambda^2_3(p_Fw)(T/\varepsilon_F)^2\ln^2(\varepsilon_F/T).
\end{equation}
In a similar fashion one can find the relaxation rate for the odd part of the imbalanced distribution. 
For this purposes one may consider a boosted frame of reference, $\varepsilon_p-pu$, 
and derive the relaxation equation for $u$ by projecting the collision integral onto the momentum mode. 
Kinematics of the respective collision is different though, and will be considered in the next section. 

To get an idea of the order of magnitude of the different timescales, it is instructive to consider the following estimates for GaAs quantum wires using experimental parameters of Ref. \cite{Barak}. For $v_F\sim 2\times10^5$ m/s and $\varkappa\sim10$, the interaction parameter is just within the applicability criterion of the perturbative expressions $g\sim1$. For the typical electron density we use $p_F\sim 10^8$ m$^{-1}$, $w\sim 10$ nm, and $\varepsilon_F\sim 1$ meV. Then for $\varepsilon\sim\varepsilon_F/4$, which is a typical excess energy of injected particles in tunneling experiments, and $T\sim0.25$K one is securely in the regime $T\ll\varepsilon^2/\varepsilon_F$. For this set of parameters $\tau^{-1}_{\text{qp}}\sim 10^{11}$ s$^{-1}$, $\tau^{-1}_{\text{qh}}\sim 10^9$ s$^{-1}$, and $\tau^{-1}_{\text{th}}\sim 10^6$ s$^{-1}$. 

\subsection{Backscattering hole mobility rates}  

Relaxation processes of low-energy excitations leading to the decay of quasiparticles near the Fermi energy do not change the numbers of right- and left-moving particles. Thus they are chirality conserving. It turns out that it is also possible to have backscattering processes. The kinematics of these collisions involves states deep in the Fermi sea, and for this reason it is useful to consider the mobility of holes 
at the bottom of the band. These processes are commonly considered from the perspective of mobile impurities in a Luttinger liquid \cite{Ogawa-Furusdaki,CastroNeto-Fisher,Imambekov-Glazman,Lamacraft,Schecter,Matveev-Andreev,Rieder}. Here we will continue 
using the kinetic equation approach for their description. The idea is then to single out hole states at the bottom of the band with small momenta, and to derive an effective kinetic equation capturing their dynamics and allowing the calculation of corresponding backscattering rates \cite{Micklitz-Rech,Polyakov,Rieder-Micklitz}.

For this purpose, let $p_1$ and $p'_1$ be momenta near the band bottom, $p_2$ and $p'_2$ lie near the right Fermi point $(+p_F)$, and $p_3$ and $p'_3$ be taken near the left Fermi point $(-p_F)$. As before, the unprimed momenta correspond to incoming states whereas primed ones are associated with outgoing states. With these conventions, we introduce the hole distribution function, $h_{p_1}=1-n_{p_1}$, and the collision integral for holes, $\St\{h_{p_1}\}=-\St\{n_{p_1}\}$. Starting from Eq. \eqref{eq:St-eee}, the latter can be cast in the form 
\begin{equation}\label{eq:St-hole}
\St\{h_{p_1}\}=\sum_{p'_1}\left[\mathcal{P}(p_1,p'_1)h_{p'_1}-\mathcal{P}(p'_1,p_1)h_{p_1}\right], 
\end{equation}
where  
\begin{equation}
\mathcal{P}(p_1,p'_1)=12\sum_{\{\sigma\}}\sum_{\stackrel{p_2p_3}{p'_2p'_3}}Wf_{p_2}(1-f_{p'_2})f_{p_3}(1-f_{p'_3})
\end{equation}
is the rate for a transition in which a hole scatters from some state $p'_1$ into $p_1$, while $\mathcal{P}(p'_1,p_1)$ denotes the  
rate for the inverse process. In the above sums, all momenta have been restricted to the discussed ranges, which explains the
combinatorial overall factor of $12$. Since both $p_1$ and $p'_1$ lie near the bottom of the band, the distribution functions $h_{p_1}$ and $h_{p'_1}$ are exponentially small $\propto e^{-\varepsilon_F/T}$ due to Pauli exclusion, and so is the collision integral of holes $\St\{h_{p}\}$. It is therefore unnecessary to account for additional exponentially small contributions in the transition rates $\mathcal{P}(p_1,p'_1)$ and $\mathcal{P}(p'_1,p_1)$, and this is why we replaced $f_{p_1}\simeq 1$ and $f_{p'_{1}}\simeq1$ in both. As in the case of the forward scattering process, the typical scale for momentum change of all three particles in a hole backscattering is set by temperature, $q_i=p'_i-p_i\sim T/v_F$. At the same time, the typical momentum of a hole is $p_1\sim\sqrt{m^*T}$ so that $q_1/p_1\sim\sqrt{T/\varepsilon_F}\ll1$. This means that the net momentum change in each scattering event is small, and holes effectively drift through the bottom of the band. Thus relaxation occurs in multiple steps and the underlying dynamics is diffusion in momentum space. Under these conditions, 
the mobile impurity falls into the universal class of problems described by Fokker-Planck equation \cite{Risken}. 
The collision integral in Eq. \eqref{eq:St-hole} can then be simplified by expanding in the small momentum step $q_1\ll p_1$, and thus maps to the differential operator 
\begin{equation}\label{eq:St-FK}
\St\{h_{p_1}\}\approx-\partial_{p_1}\left[\mathcal{A}(p_1)h_{p_1}\right]+\frac{1}{2}\partial^2_{p_1}\left[\mathcal{B}(p_1)h_{p_1}\right].
\end{equation}
Here we introduced 
\begin{equation}
\mathcal{A}(p_1)=-\sum_{q_1}q_1\mathcal{P}_{q_1}(p_1),\quad \mathcal{B}(p_1)=\sum_{q_1}q^2_1\mathcal{P}_{q_1}(p_1),
\end{equation}
and used the short-hand notation $\mathcal{P}_{q_1}(p_1)=\mathcal{P}(p'_1,p_1)$. The diffusion coefficient in momentum space $\mathcal{B}(p_1)$ is a function of the hole-momentum $p_1$ varying  on a scale set by $p_F$. For holes at the bottom of the band, one may thus approximate  $\mathcal{B}(p_1)$ by its value at $p_1= 0$, in the following simply denoted by $\mathcal{B}$ without argument.
Furthermore, the drift coefficient $\mathcal{A}(p_1)$ is readily obtained from noting that the collision integral \eqref{eq:St-FK} has to vanish  
for hole distributions of an equilibrium Boltzmann form. This condition leads to the relation $\mathcal{A}(p)=p\mathcal{B}/2m^*T$. 
 
The rest of the calculation depends on the structure of the amplitude for a given kinematics of the three-particle process. In calculating $A$ from Eq. \eqref{eq:A} for the momentum configuration under consideration, and up to small corrections in $T/\varepsilon_F\ll1$, it is sufficient to approximate $p_1\approx0$, $p_2\approx +p_F$ and $p_3\approx-p_F$. Momentum and energy conservations provide additional restrictions on the transferred momenta, enforcing that $q_2\approx q_3\approx -q_1/2$, again up to small corrections in $T/\varepsilon_F\ll1$. As a result, the amplitude $A$ can be parametrized only by a single momentum $q_1$. Expanding Eq. \eqref{eq:A} and summing over spins one then finds 
\begin{equation}\label{eq:A-bs-Spin}
\sum_{\{\sigma\}}|A|^2=\frac{6}{\varepsilon^2_FL^4}V^2_{p_F}(V_{p_F}-V_{2p_F})^2\frac{p^2_F}{q^2_1}. 
\end{equation}  
The singularity of $A$ at small momenta is cancelled in the spinless case. The underlying reason is exactly the same as discussed above in the context of quasiparticle thermalization processes. Specifically, for the long-range interaction model with Eq. \eqref{eq:V-C} one finds   
\begin{equation}\label{eq:A-bs-C}
A\approx\frac{9}{16\varepsilon^2_FL^4}\left(\frac{2e^2}{\varkappa}\right)^4\lambda^2_1(p_Fw)\ln^2\left(\frac{p_F}{|q_1|}\right), 
\end{equation}
whereas for the screened model 
\begin{equation}\label{eq:A-bs-SC}
A\approx\frac{9(\ln4-1)^2}{\varepsilon^2_FL^4}\left(\frac{2e^2}{\varkappa}\right)^4\lambda^2_2(p_Fd). 
\end{equation}
In order to perform the remaining momentum integrations implicit in the definition of $\mathcal{B}$, one can approximate delta functions in the scattering probability by $\delta_{P,P'}\delta(E-E')\approx\frac{1}{v_F}\delta(q_2-q_3)\delta_{q_2,-q_1/2}$. This removes two integrations out of five, and gives 
\begin{equation}
\mathcal{B}=\frac{12L}{v_F}\!\!\!\sum_{q_1k_2k_3}\!\!q^2_1\sum_{\{\sigma\}}|A|^2f_{k_2-\frac{q_1}{2}}(1-f_{k_2})f_{k_3+\frac{q_1}{2}}(1-f_{k_3}),
\end{equation} 
where we shifted momenta $p_{2,3}$ to the respective Fermi points, $\pm p_F+k_{2,3}$, and linearized 
the dispersion relation in all Fermi occupation functions.  Finally, using 
the tabulated integral
\begin{equation}
\sum_{k}f_{k+q}(1-f_k)=\frac{L}{2\pi}qb_q, \quad b_q=\frac{1}{e^{v_Fq/T}-1},
\end{equation}
where $b_q$ is the equilibrium Bose distribution, we arrive at the general expression
\begin{equation}\label{eq:B}
\mathcal{B}=\frac{6\pi}{v_F}\left(\frac{L}{2\pi}\right)^3\sum_{q_1}q^4_1\sum_{\{\sigma\}}|A|^2b_{q_1/2}\big(1+b_{q_1/2}\big). 
\end{equation}    
A notable feature of this expression is that it is entirely expressed in terms of bosonic modes. In essence, this is a manifestation of bosonization at the level of fermionic kinetic theory, as the occupation of an electron-hole pair near one of the Fermi points integrated over the center of mass momentum is equivalent to a collective boson emitted/absorbed in a course of hole diffusion. It will be shown in the subsequent section that structurally the same expression for $\mathcal{B}$ can be obtained from a purely bosonic formulation of the problem. Finally, inserting Eq. \eqref{eq:A-bs-Spin} into Eq. \eqref{eq:B} one finds the momentum space diffusion coefficient of spin-$1/2$ holes
 \begin{equation}
 \mathcal{B}=\frac{768\ln^2(2)}{\pi}g^4\lambda^2_3(p_Fw)\left(\frac{T}{\varepsilon_F}\right)^3p^2_F\varepsilon_F.
 \end{equation}
The corresponding backscattering relaxation rate can be found from Einstein relation adopted to diffusion in momentum space, $\Delta p^2=\mathcal{B}\tau_{\text{dh}}$. The notation $\tau_{\text{dh}}$ is meant to emphasize kinetics of a deep hole as opposed to earlier notation $\tau_{\text{qh}}$ describing quasiholes near Fermi energy. Thus for $\Delta p^2\simeq m^*T$ the result is (omitting numerical factosr for brevity)
 \begin{equation}
 \tau^{-1}_{\text{dh}}\simeq g^4\lambda^2_3(p_Fw)(T/\varepsilon_F)^2.
 \end{equation} 
Finally we recall that the mobility of particles $\varsigma$ is related to the diffusion constant by the simple kinetic formula
$\varsigma=T/\mathcal{B}$, and therefore $\varsigma\propto 1/T^2$. 

The result is different in the spinless case. From Eqs. \eqref{eq:A-bs-C}, \eqref{eq:A-bs-SC} and \eqref{eq:B} one finds $\mathcal{B}\propto T^5$ in both models, modulo a logarithmic factor $\ln^2 T$ in the Coulomb case, and thus $\tau^{-1}_{\text{dh}}\propto T^4$ and $\varsigma\propto 1/T^4$. The results discussed in this section are again perturbative in the interaction. The power laws in the temperature dependence of relaxation rates are, however, generic and  also apply to the strongly interacting regime, as we further elaborate below, see also Refs. \cite{Schecter,Matveev-Andreev}.  
  
\subsection{Electron-phonon relaxation rates}\label{sec:el-ph}

Apart from the purely electronic mechanisms of relaxation electrons may scatter on phonons, disorder, and sample imperfections thus relaxing their energy and momentum. At extremely low temperatures phonons are not expected to be efficient at cooling the electronic sub-system. On the other hand, electron-phonon scattering has no such severe phase space restrictions like the three-particle collisions considered above. It is thus instructive to estimate the temperature dependence for the corresponding relaxation rate. Unlike the previous studies of electron-phonon relaxation in multichannel quantum wires \cite{Gurevich,Sergeev}, and phonon-induced backscattering relaxation \cite{Seeling-1,Seeling-2}, we focus on the complementary effect of soft collisions in a single-channel geometry of strictly 1D electrons and 3D phonons.    

The coupling of electrons and phonons is described by the collision integral \cite{ALJS-ElPh}
\begin{align}\label{eq:St-ep}
&\St\{n_p,N_q\}=\nonumber \\ 
&\sum_{p'q}W_-[n_{p'}(1-n_{p})N_q-n_{p}(1-n_{p'})(1+N_q)]\nonumber \\ 
&+\sum_{p'q}W_{+}[n_{p'}(1-n_{p})(1+N_q)-n_{p}(1-n_{p'})N_q]\,,
\end{align}
where the scattering rate 
\begin{equation}
W_{\pm}(p,p',q)=(2\pi)|A(q)|^2
\delta(\varepsilon_p-\varepsilon_{p'}\pm\omega_q)\delta_{p=p'\pm q_x}
\end{equation}
describes phonon emission and absorption processes with an amplitude $A(q)=\sqrt{\frac{1}{2\varrho \mathcal{V}\omega_q}}(D|q|+i\Lambda)$.  
Here we took into account that at the level of the leading Born approximation, the probabilities of scattering for direct and reverse processes are the same. In the amplitude we include both deformation $(D)$ and piezoelectric $(\Lambda)$ couplings, $\varrho$ is the mass density, $q_x$ the phonon wave-vector along the wire, and $\mathcal{V}$ is the system volume. For simplicity we assume only a single acoustic branch $\omega_q=s|q|$, with sound velocity $s$. 

For equilibrium Fermi and Bose distribution functions of electrons and phonons respectively, $n_p\to f_p$ and $N_q\to b_q$,  
the collision integral in Eq. \eqref{eq:St-ep} vanishes due to detailed balance condition. As in the above example of the distribution imbalance relaxation, we  then assume that electrons are hot. That is,  at an excess temperature $T+\Delta T$ with respect to the temperature $T$ of 
lattice phonons. Electron-phonon collisions tend to relax $\Delta T$, and the corresponding rate for relaxation 
can be found by projecting the collision integral onto the energy mode, $\sum_p\epsilon_p\dot{n}_p=-\sum_p\epsilon_p\St\{n_p,N_q\}$, with $\epsilon_p=\varepsilon_p-\varepsilon_F$.
To linear order in $\Delta T$ one finds from the phonon emission processes of hot electrons, $\partial_t\Delta T=-\Delta T/\tau_{\mathrm{ep}}$, where
\begin{equation}
\tau^{-1}_{\mathrm{ep}}=-\frac{6v_F}{\pi T^3L}\sum_{pp'q}W
\epsilon_p\omega_q f_{\varepsilon_p}(1-f_{\varepsilon_p})
(f_{\varepsilon_p+\omega_q}+b_{\omega_q})\,.
\end{equation}
Upon completion of the remaining momentum integrations, we then find to leading order in $T$
\begin{equation}\label{eq:tau-ep}
\tau^{-1}_{\mathrm{ep}}=\frac{9\zeta(3)}{8\pi^3}T(\Lambda^2/s^2v_F\varrho).
\end{equation}
The scattering rate due to the deformation potential is parametrically  weaker, scaling as $\tau^{-1}_{\mathrm{ep}}\propto T^3$. The backscattering mechanism results in an activated temperature dependence $\propto e^{-T_A/T}$ with $T_A=2sp_F$. It is straightforward to generalize Eq. \eqref{eq:tau-ep} to the case when electronic relaxation occurs via several acoustic branches. Notice also that the piezoelectric potential may have complicated angular dependence in case of wires oriented arbitrarily with respect to the crystallographical axis of the sample. A proper angular averaging would change then numerical factors in Eq.~\eqref{eq:tau-ep} where we took the simplest geometry. Luttinger liquid effects lead to renormalization of the linear-$T$ behavior and transform it into a power-law with interaction dependent exponent $\propto T^K$, where $K=v_F/u$ is the ratio of Fermi and plasmon velocities. In the TL model $u=v_F\sqrt{1+V_0/\pi v_F}$. 

\subsection{Spin-charge scattering rates}\label{sec:spin-charge}

The applicability of the Born approximation, used to construct the quantum amplitude for triple particle processes captured by Eq. \eqref{eq:A}, requires that incoming spin-$1/2$ quasiparticles have sufficiently high energy compared to the typical scale of interparticle interaction $\varepsilon\gg m^*v_FV_0$. 

In the generic interacting environment of a 1D quantum fluid, quasiparticle excitations break down into spin and charge modes. At the level of linear Luttinger liquid theory, spin-charge separation is an exact property of the model \cite{Giamarchi}. At weak coupling, the splitting between velocities of collective spin $(v_\sigma)$ and charge $(v_\rho)$ density waves is related to the forward scattering component of the interaction $v_\rho-v_\sigma\sim V_0$ (recall that for repulsive interactions $v_\rho>v_\sigma$). Assuming then thermal excitations with $\varepsilon\sim T$, the Born condition can be equivalently formulated as $T/(m^*v_F)\gg v_\rho-v_\sigma$. In other words, for fermionic quasiparticles to preserve their integrity the excitation energy (or temperature) should be bigger than the energy scale for spin-charge separation. 

The interplay of spectrum nonlinearities and interactions leads to spin-charge coupling \cite{Brazovskii,Nayak}. Although irrelevant in the renormalization group sense, the newly emerging higher order operators capture the attenuation of quasiparticles. The kinetic properties of 1D quantum liquids with spin-charge coupling are not fully understood. There are basically two possible approaches one may pursue. The first is to refermionize the nonlinear bosonic theory to obtain an effective description in terms of dressed quasiparticles: holons and spinons. Holon relaxation was considered in Refs. \cite{Schmidt,Pereira-Sela} based on non-Abelian bosonization \cite{GNT}. The advantage of this complex theory is that, in principle, it allows to go beyond the weakly interacting limit for spinful fermions. Alternatively, one may choose to continue working in the bosonic language. In the limit of weak backscattering one can then account for spin-charge interaction perturbatively in the basis of well-defined spin and charge modes. This second procedure is limited to weak interactions $V_{2p_F}\ll V_0\ll v_F$. To complement previous studies, we follow in this section the second path. In part this will enable us to explore the fermion-boson duality. We delegate technical details of bosonization to the Appendix and elucidate here the impact of spin-charge scattering on various decay rates.  

The lowest order nonlinearity, compatible with SU(2) symmetry of the problem is cubic. It contains one charge and two spin operators. Treating this term in a perturbative Golden rule expansion generates a collision kernel that describes the decay of a plasmon into two spin modes
 $\rho\to\sigma\sigma$. It reads   
\begin{align}\label{eq:St-css}
&\St\{N^\rho,N^\sigma\}=\nonumber \\ 
&-\sum_{q_1q_2}W\left[N^\rho_q(1+N^\sigma_{q_1})(1+N^\sigma_{q_2})-(1+N^\rho_q)N^\sigma_{q_1}N^\sigma_{q_2}\right],
\end{align}
where $N^{\rho/\sigma}$ are the bosonic occupations of charge ($\rho$) and spin ($\sigma$) excitations. The scattering probability  
\begin{equation}
W=2\pi|A|^2\delta_{q=q_1+q_2}\delta(\omega^\rho_q-\omega^\sigma_{q_1}-\omega^\sigma_{q_2})
\end{equation}
contains an amplitude scaling cubically with momenta of the bosons $|A|^2=(\pi^3/8L)|q||q_1||q_2|\Gamma^2_{\rho\sigma\sigma}$. The perturbative result for the coupling constant is $\Gamma_{\rho\sigma\sigma}=V'_{2p_F}/\sqrt{2}\pi^2$, where the prime denotes the derivative with respect to $p_F$. Note that it thus vanishes for the integrable case of constant interaction.
At smallest momenta the dispersion relations are linear $\omega^{\rho/\sigma}=v_{\rho/\sigma}|q|$. The kinematics of this process uniquely fixes momenta in the final state. Indeed, for concreteness let $q>0$, then $q_1=q(v_\rho+v_\sigma)/2v_\sigma$ and $q_2=-q(v_\rho-v_\sigma)/2v_\sigma$, which means that spin waves are counterpropagating. From dimensional analysis it 
becomes apparent that $\St\{N^\rho,N^\sigma\}$ defines the decay rate of a plasmon, and one can introduce the characteristic rate $\tau^{-1}_{\rho}=\sum_{q_1q_2}W\simeq q^3(V'_{2p_F})^2(v^2_\rho-v^2_\sigma)/v^3_\sigma$. For the sake of an estimate, one may now take $V'_{2p_F}\sim V_{2p_F}/p_F$ and replace $v_{\rho/\sigma}\sim v_F$, except in their difference where 
$v_\rho-v_\sigma\sim V_0$, and finds the life-time 
\begin{equation}
\tau^{-1}_{\rho}\sim \varepsilon_F(V_0/v_F)(V_{2p_F}/v_F)^2(q/p_F)^3. 
\end{equation}  
Notice the nonanalytic dependence of interaction $\propto V^3$. For thermal plasmons the relaxation rate can be calculated from Eq. \eqref{eq:St-css} by a projection onto an energy mode. We observe that as $|q_2|\ll q$ the relaxation occurs by small energy transfer from right-movers to left-movers (or vise-versa) so that interbrach processes are slow. Assuming that right-moving excitations are hotter by $\Delta T$, and in complete analogy to the fermionic case, we find 
\begin{equation}
\sum_{q>0}\omega^\rho_q\partial_t N^\rho_q=\sum_{q>0}\omega^\rho_q\St\{N^\rho,N^\sigma\}.
\end{equation} 
The left-hand-side is straightforwardly evaluated, noting that 
\begin{equation}
\partial_tN^\rho_q=\partial_TN^\rho_q\partial_t\Delta T=\frac{\omega^\rho_q\partial_t\Delta T}{4T^2\sinh^2(\omega^\rho_q/2T)},
\end{equation}
which after momentum integration gives a factor of $(\pi LT/6v_\rho)\partial_t\Delta T$. The right-hand-side can be linearized with the usual substitution $N_q=b_q+b_q(1+b_q)\phi_q$, where $\phi_q=\omega_q\Delta T/T^2$ for the case of a thermal imbalance. After some algebra one finds 
\begin{align}
&\sum_{q>0}\omega^\rho_q\St\{N^\rho,N^\sigma\}=\nonumber \\ 
&-\Delta T\sum_{qq_1q_2}(\omega^\rho_q/T)^2W(1+b^\rho_q)b^\sigma_{q_1}b^\sigma_{q_2},
\end{align}
where we repeatedly used energy conservation and the detailed balance condition. Performing the final integrations, we then arrive at  
\begin{equation}\label{eq:tau-css}
\tau^{-1}_{\rho}=\frac{3\pi}{16}\Gamma^2_{\rho\sigma\sigma}(T^3/v^4_\sigma)F(v_\sigma/v_\rho),
\end{equation}
where the dimensionless function reads
\begin{equation}
F(x)=x(1-x^2)\int^{\infty}_{0}z^5(1+b_z)b_{z_+}b_{z_-} dz
\end{equation}
with $b_z=(e^z-1)^{-1}$ and $z_\pm=z(1\pm x)/2$. One can readily check that $F\to32\pi^4/15$ in the limit $x\to1$. 

The same scattering process can be alternatively viewed as a mutual spin-charge friction. 
Physically, this is analogous to the electron-phonon drag effect, typically studied in the context of thermoelectricity, 
or Coulomb drag in double-layers \cite{Drag-Review} and spin Coulomb drag \cite{Vignale}.  
In each of these examples momentum transfer between interactively coupled systems leads to dragging of one sub-system by the flow of the other. For instance, in the context of spin physics in Luttinger liquids, generation of spin current is possible by Coulomb drag \cite{Pustilnik}. To estimate the spin-charge drag rate, one can consider a boosted frame of reference for spin and charge excitations with mismatched boost velocities $u_{\rho/\sigma}$. The scattering leads to momentum exchange between spins and charge and, as a result, to relaxation $\partial_t u_\rho=-(u_\rho-u_\sigma)/\tau_{\rho\sigma}$. To capture this effect, we linearize the collision integral for $N^{\rho/\sigma}(\omega_q-uq)$ with respect to $u$, for both spin and charge occupations, and then project onto the momentum mode to calculate the rate of momentum loss by (e.g.) charge modes
\begin{align}
&\partial_tP_\rho=\sum_qq\St\{N^\rho,N^\sigma\}=\nonumber \\ 
&-(\Delta u/T)\sum_{qq_1q_2}q^2W(1+b^\rho_q)b^\sigma_{q_1}b^\sigma_{q_2}. 
\end{align}  
When we compare this to $\partial_tP_\rho=(\pi LT^2/3v^3_\rho)\partial_tu_\rho$, we find that thus defined drag relaxation rate $\tau^{-1}_{\rho\sigma}$ coincides with Eq. \eqref{eq:tau-css} up to a constant factor. It is perhaps useful to note that $\tau^{-1}_{\rho\sigma}\propto T^3$ is consistent with the expectation that Coulomb drag transresistivity between double quantum wires due to interwire momentum transfer from spin-charge coupling at zero magnetic field scales as $\rho_\text{drag}\propto T^5$ \cite{Pereira-Sela}. Indeed, this rate is accompanied by two thermal phase space factors $\sim T$ per wire, thus leading to $T^5$. In the drag problem, the factor $q^3$ results from the width of the dynamic charge structure factor and the underlying scattering that gives rise to its $\propto q^3$ width is precisely the decay of a charge boson into two spin bosons.   
   
The next in complexity is a quartic nonlinearity in spin-charge coupling which leads to two-boson scattering $\rho\sigma\to\rho\sigma$. In particular, we consider backscattering of spin excitations on plasmons. Such scattering processes correspond to the diffusion of spin excitations near the spectral edge, and the goal is to calculate the corresponding diffusion constant. As alluded to earlier, the discussion parallels the previous calculation of the backscattering of a deep hole in the fermionic language. 
The corresponding collision integral reads 
\begin{align}\label{eq:St-scsc}
\St\{N^\rho,N^\sigma\}=-\!\!\!\sum_{q_2q'_1q'_2}\!\!W \left[N^\sigma_{q_1}(1+N^\sigma_{q'_1})N^\rho_{q_2}(1+N^\rho_{q'_2})\right. \nonumber \\ 
-\left.N^\sigma_{q'_1}(1+N^\sigma_{q_1})N^\rho_{q'_2}(1+N^\rho_{q_2})\right]. 
\end{align}     
The scattering rate for this process is given by 
\begin{equation}
W=2\pi|A|^2\delta_{Q,Q'}\delta(E-E')
\end{equation}
with the amplitude $|A|^2=(\Gamma_{\rho\sigma}/8L)^2|q_1q'_1q_2q'_2|$, where the coupling constant at the perturbative level 
 reads $\Gamma_{\rho\sigma}=V''_{2p_F}$. The notations for momentum and energy conservation here are $Q=q_1+q_2$  and $E=\omega^\sigma_{q_1}+\omega^\rho_{q_2}$. Let momenta $q_1$ and $q'_1$ correspond to the initial and final states of the spin excitation near the spectral edge. Kinematically each momentum is of the order of the Fermi momentum, $q_1\sim q'_1\sim p_F$, while their difference, $q'_1 - q_1\sim T/v_F$, is small. This corresponds to a small momentum change in each collision, which is accompanied by the excitation of plasmons at low momenta $q_2\sim q'_2\sim T/v_F$. For this reason the low-energy description based on Eq. \eqref{eq:St-scsc} is sufficient to capture this physics. Under the specified conditions and in complete analogy with the fermionic case, we can convert  the collision integral into a Fokker-Planck differential operator, thus describing the diffusion of spins. Indeed, for momenta $q\sim p_F$ the occupation is small, $N_{q}^\sigma\propto e^{-\varepsilon_\sigma/T}\ll1$, and correspondingly $1+N^\sigma\approx1$, where $\varepsilon_\sigma$ is the band width of spin excitations. The latter is parametrically of the order of the spin exchange coupling.
Viewing Eq. \eqref{eq:St-scsc} as the collision integral for spins, we thus write 
\begin{equation}
\mathrm{St}\{N^\sigma_{q_1}\}=\sum_{q'_1}[\mathcal{P}(q_1,q'_1)N^\sigma_{q'_1}-\mathcal{P}(q'_1,q_1)N^\sigma_{q_1}],
\end{equation}
where
\begin{equation}
\mathcal{P}(q'_1,q_1)=\sum_{q_2q'_2}W N^\rho_{q'_2}(1+N^\rho_{q_2}),
\end{equation} 
is the transition rate for spin scattering processes. More specifically, it describes a collision with momentum transfer $\delta q$, in which a spin is scattered out of the initial state $q_1$. It can thus be rewritten as $\mathcal{P}(q'_1,q_1) = \mathcal{P}_{\delta q}(q_1)$, and
following the same prescription, the transition rate for the inverse process reads $\mathcal{P}(q_1, q'_1) = \mathcal{P}_{-\delta q}(q_1 +\delta q)$. Performing then a small-momentum expansion,
\begin{align}
&\mathcal{P}(q_1,q'_1)N^\sigma_{q'_1}\approx \mathcal{P}_{-\delta q}(q_1)N^\sigma_{q_1}\nonumber \\ 
&+\delta q\partial_{q_1}[\mathcal{P}_{-\delta q}(q_1)N^\sigma_{q_1}]+\frac{\delta q^2}{2}\partial^2_{q_1}[\mathcal{P}_{-\delta q}(q_1)N^\sigma_{q_1}],
\end{align}
the collision integral of spin excitations takes the simplified form
\begin{equation}
\mathrm{St}\{N^\sigma_{q_1}\}=-\partial_{q_1}[\mathcal{A}_{\rho\sigma}(q_1)N^\sigma_{q_1}]+\frac{1}{2}\partial^2_{q_1}[\mathcal{B}_{\rho\sigma}(q_1)N^\sigma_{q_1}],
\end{equation}
where 
\begin{equation}
\mathcal{A}_{\rho\sigma}=-\sum_{\delta q}\delta q\mathcal{P}_{\delta q}(q_1),\quad
\mathcal{B}_{\rho\sigma}=\sum_{\delta q}\delta q^2\mathcal{P}_{\delta q}(q_1).
\end{equation}
At this stage we focus on the derivation of $\mathcal{P}_{\delta q}(q_1)$. The momentum conservation implicit in $W$ removes the $q'_2$ integration.  We then notice that distribution functions limit the typical momentum transfer and momenta of plasmons to $q_2\sim\delta q\sim T/v_F$. At the same time, the typical momentum of spins at the spectral edge is $q_1\sim p_F$ and it is sufficient to calculate $\mathcal{P}_{\delta q}(p_F)$. With these observations at hand, we can now approximate energy conservation by $\delta(E-E')\approx\frac{1}{v_\rho}\delta(q_2-\delta q/2)$. This removes the $q_2$ integral, and  we thus arrive at 
\begin{equation}\label{eq:P-scsc} 
\mathcal{P}_{\delta q}=\frac{V^2_{\rho\sigma}}{1024Lv_\rho}\frac{(\delta q/p_F)^2}{\sinh^2(v_\rho\delta q/4T)},
\end{equation}  
with the notation $V_{\rho\sigma}=p^2_F\Gamma_{\rho\sigma}$. Finally, this defines the diffusion coefficient of spins in momentum space  
associated to $\rho\sigma\to\rho\sigma$ scattering channel
\begin{equation}
\mathcal{B}_{\rho\sigma}=\frac{\pi^3}{30}(V_{\rho\sigma}/v_\rho)^2(T/p_Fv_\rho)^5p^3_Fv_\rho. 
\end{equation}
Exactly the same result for the temperature dependence of the momentum space diffusion constant pertains to scattering in the charge channel, proceeding via the $\rho\rho\to\rho\rho$ scattering process. With the replacement of the respective coupling constant one concludes that $\mathcal{B}_{\rho\rho}\propto T^5$. 

In addition to spin-charge scattering, nonlinearities also allow for spin-spin scattering. Importantly for the momentum space diffusion, scattering processes with spin-flips are enhanced. They are thus described by a different scaling of the probability with momentum as compared to Eq. \eqref{eq:P-scsc}. That is,
\begin{equation}\label{eq:P-ssss} 
\mathcal{P}_{\delta q}=\frac{V^2_{\sigma\sigma}}{8\pi^2Lv_\sigma}\frac{1}{\sinh^2(v_\sigma\delta q/4T)},
\end{equation}  
and this crucial detail is technically speaking traced back to the non-commutativity of spin operators when calculating the corresponding amplitude. The importance of spin flips is also apparent at the level of fermions. Indeed, the ratio of scattering rates between spinless and spinful cases has exactly the same parameter $(q/p_F)^2\ll1$ as the ratio between probabilities in Eqs. \eqref{eq:P-scsc} and \eqref{eq:P-ssss}. The resulting diffusion constant in the spin-spin channel is then     
\begin{equation}
\mathcal{B}_{\sigma\sigma}=\frac{4\pi}{15}(V_{\sigma\sigma}/v_\sigma)^2(T/p_Fv_\sigma)^3p^3_Fv_\sigma.
\end{equation}
A microscopic calculation of the respective coupling constants for the different scattering channels is a challenging task. Known approaches include weak coupling results obtained via mobile impurity model \cite{Rieder}, results for Kondo polarons \cite{Lamacraft}, and calculations in the strong interaction limit within the non-Abelian bosonization framework \cite{Pereira-Sela}, as well as a model departing from the Wigner crystal limit \cite{Klironomos}. 

Two-boson processes also contribute to the thermalization rates \cite{Protopopov,Lin,Apostolov}. For charge excitations this results in a subleading correction to Eq. \eqref{eq:tau-css}. In the spin sector the situation is, however, different since at the cubic level of nonlinearities spins are kinematically forbidden to scatter. In both cases nonlinearity of the bosonic spectrum plays an important role to open phase space for such collisions. In order to generalize the present model, consider  first the charge sector and assume a weakly anharmonic dispersion of plasmons, $\omega^\rho_q\approx v_\rho |q|(1-(\ell q)^2)$. Assume now that a right-moving boson with momentum $q_1\gtrsim T/v_\rho$ is injected into the Luttinger liquid. For this setting the collision term from Eq. \eqref{eq:St-scsc}, with
replacement $N^\sigma\to N^\rho$, dictates that the dominant process limiting the lifetime of the injected boson is due to 
scattering with inter-branch momentum transfer. Indeed, for $q_1,q_2,q'_1>0$ momentum conservation implies that $q'_2$ is order $q^3$, since energy conservation fixes $q'_2\approx -\frac{3\ell^2}{2}q_1q_2q'_1$. Curiously, even though finite dispersion curvature parameter $\ell$ is crucial to resolve the kinematic constraints it drops out from the corresponding rate provided that $q_1\ll \sqrt[3]{T/v_\rho\ell^2}$. In this regime $v_\rho |q'_2|\ll T$, implying that $N^\rho_{q'_2}\approx T/\omega^\rho_{q'_2}$  and $q'_2$ cancels out from $W$. The decay rate  then scales parametrically as $\tau^{-1}_{\rho}\propto Tq^4$, which is applicable as long as $T/v_\rho\lesssim q\ll\sqrt[3]{T/v_\rho\ell^2}$.

For thermal plasmons, this rate can be estimated more accurately by projecting the collision integral onto the energy mode. Assuming that the boson with momentum $q_1$ is more energetic, namely hotter by a temperature difference $\Delta T$,
one finds upon repeating the steps from the previous similar calculations  
\begin{equation}
\tau^{-1}_{\rho}=\frac{6v_\rho}{\pi LT}
\sum_{\stackrel{q_1q_2}{q'_1q'_2}}
\frac{\omega_{q_1}\omega_{q_2}}{T^2}WN^\rho_{q_1}N^\rho_{q_2}(1+N^\rho_{q'_1})(1+N^\rho_{q'_2}).
\end{equation}
For the kinematics of the process specified above, one sum is removed by momentum conservation setting $q'_1=(q_1+q_2)$. 
Energy conservation removes another integral, setting $q'_2=-\frac{3\ell^2}{2}q_1q_2(q_1+q_2)$. 
The remaining integrals can, after rescaling of momentum variables in units of temperature, 
 be brought to a dimensionless double-integral.   
This results in 
\begin{equation}
\tau^{-1}_{\rho}=\frac{3c_{12}}{(4\pi)^4}(V_{\rho\rho}/v_\rho)^2T(T/p_Fv_\rho)^4,
\end{equation}
where the coefficient $c_{12}=\int^{\infty}_{0}\frac{x^2y^2(x+y)e^{x+y}dxdy}{(e^x-1)(e^y-1)(e^{x+y}-1)}$ and $V_{\rho\rho}=p^2_F\Gamma_{\rho\rho}$. 

Two-spin scattering can be analyzed in the same way, starting out from Eq. \eqref{eq:St-scsc} by changing $N^\rho\to N^\sigma$. The crucial difference is in the momentum dependence of the scattering rate, which is enhanced by spin-flip processes. The resulting spin wave thermalization rate due to two-boson scattering processes reads 
\begin{equation}
\tau^{-1}_{\sigma}\sim(V_{\sigma\sigma}/v_\sigma)^2T(T/p_Fv_\sigma)^2. 
\end{equation}
This final estimate exhausts all possible scattering processes emerging from the quartic corrections to the 
linear Luttinger liquid model.


\section{Transport properties}\label{sec:transport}

Thermalization and equilibration processes bear consequences on the temperature dependence of kinetic coefficients in Fermi-Luttinger liquids. Before turning to their discussion we open by briefly recapitulating the electrical and thermal transport properties predicted by linear Luttinger liquid theory. For a more expanded review see e.g. Refs. \cite{Maslov,Fisher-Glazman}. 

For a clean single-mode quantum wire, adiabatically connected to Fermi liquid leads, linear Luttinger liquid theory predicts that the conductance $\mathcal{G}$ remains unrenormalized by interactions. At zero temperature the conductance thus takes the same value as for noninteracting electrons $\mathcal{G}_0=e^2/\pi$ \cite{Glazman,Maslov-Stone,Safi-Schulz,Ponomarenko}. In such a two-terminal setup, finite resistance results only from scattering of electrons from the boundary regions, where the wide Fermi liquid reservoirs are connected to the narrow quantum wire. This result holds also for smoothly inhomogeneous wires as long as the correlation radius $\xi$ of 
the disorder potential is sufficiently large and electron backscattering processes are exponentially suppressed in $e^{-2\xi p_F}\ll1$.   
As we discuss below, equilibration processes in nonlinear Fermi-Luttinger liquids lead to finite temperature corrections 
to the quantized conductance. Since the interaction-induced equilibration occurs via backscattering of holes near the bottom of the band, these corrections are of the activated form $\delta\mathcal{G}\propto e^{-\varepsilon_F/T}$ in short wires \cite{Lunde,LRM,Rieder-Micklitz}. For sufficiently long wires, they become however more pronounced scaling then as $\delta \mathcal{G}\propto T^2$ in both, weakly and strongly interacting regimes \cite{Micklitz-Rech,Matveev-Andreev-LLG}.   

The physical picture of heat transport in linear Luttinger-liquids is different. Electrons entering the wire from one of the leads break up into plasmons, which then propagate along the wire as in a waveguide, carrying the energy current. In contrast to electrons, plasmons experience backscattering at the ends of the wire due to mismatch of the plasmon velocity in the wire and the Fermi velocity in the leads. 
This leads in essence to an analog of the Kapitza boundary thermal resistance. 

Multiple plasmon back reflections from both ends of the wire give further rise to   interference like in a Fabry-P\'erot interferometer.  
Transmission coefficients for plasmons through the wire are thus described by a Fresnel type formula, and the thermal conductance reads 
$\mathcal{K}=2g\mathcal{K}_0/(g^2+1)$ \cite{Kane-Fisher,Fazio}, where $\mathcal{K}_0=\pi T/3$ is the thermal conductance quantum. 
As a consequence, the Lorentz ratio $\mathcal{L}=\mathcal{K}/T\mathcal{G}$ in a linear Luttinger liquid takes a non-universal and interaction dependent value $\mathcal{L}/\mathcal{L}_0=2g/(g^2+1)<1$, with $\mathcal{L}_0=\pi^2/3e^2$. This also implies violation of the
Wiedemann-Franz law.  
 
Going beyond the linear Luttinger liquid description, thermalization processes enable energy exchange between right- and left-moving excitations. This results in temperature dependent corrections to the thermal conductance, $\delta \mathcal{K}\propto T^3$ \cite{AL-K}, for wires of intermediate length for which backscattering of deep holes is still negligible. For long wires, equilibration processes prevail and suppress thermal conductance as $\mathcal{K}\propto 1/L$, now vanishing inversely proportional with the wire length. The 
thermal conductivity, however, remains well defined and exhibits a complex temperature dependence \cite{AL-RKS,Matveev-Ristivojevic,Samanta}.
       
We next present the above predictions for Fermi-Luttinger liquids in a more detail fashion, and discuss other related transport properties.       

\subsection{Electrical response}\label{sec:electrical}

Consider a quantum wire of length $L$ adiabatically connected to leads of non-interacting electrons, and biased by a finite voltage $V$. 
As particles enter the wire, they initially keep memory of the lead they departed from and specifically of the respective chemical potential. Deeper inside the wire, electrons interact and experience backscattering, equilibrating thus the difference in chemical potentials 
of left- and right-moving electrons. Particle number conservation then implies a simple relation between the applied voltage $V$, current $I$ carried by the wire, and rate for backscattering
\begin{equation}
\label{eq:Landauer-G}
\mathcal{G}_0V=I-e\dot{N}^R,
\end{equation}
where $\dot{N}^R\equiv\partial_tN^R$ here is the rate of change in the number of right-moving particles. In the absence of  
backscattering, $\dot{N}^R=0$, and Eq.~\eqref{eq:Landauer-G} reduces to the Landauer conductance of a perfectly transmitting channel. 
To make further progress, we notice that the current inside the wire comprises of two terms, 
\begin{equation}\label{eq:I-mu-u}
I=2e\frac{\Delta\mu}{2\pi}+enu,
\end{equation} 
resulting from the imbalance in the chemical potentials of left- and right-moving electrons, $\Delta\mu=\mu^R-\mu^L$,  
respectively, the drift of the electron flow with velocity $u$. Each contribution may individually vary along the wire, and current conservation implies that this cannot happen in an independent fashion. In fact, both $\Delta\mu$ and $u$ can be expressed in terms of the backscattering rate $\dot{N}^R$, and Eq. \eqref{eq:Landauer-G} then defines the conductance of interacting electrons.

To calculate $\dot{N}^R$ microscopically one has to integrate the right-hand-side of the Boltzmann equation over all positive momenta $\dot{N^R}=\sum_{p>0}\St\{n_p\}$. For the kinematics of the leading backscattering process in short wires (to be specified momentarily), the collision integral can be approximated by the Fokker-Planck kernel for holes. From Eq. \eqref{eq:St-FK} it then follows that 
\begin{equation}
\dot{N}^R=
2\sum_{p}\left[\partial_p(\mathcal{A}h_p)-\frac{1}{2}\partial^2_p(\mathcal{B}h_p)\right]=\frac{L\mathcal{B}}{2\pi}\left(\partial_ph_p\right)_{p=0}.
\end{equation}
To determine the discontinuity in the derivative of the hole's distribution function, one thus has to solve the Fokker-Planck equation, 
\begin{equation}
\partial_p\left[-\frac{p}{m^*T}h_p+\partial_ph_p\right]=0,
\end{equation}
supplemented by the boundary conditions $h_p=e^{p^2/2m^*T}e^{-\mu^{R/L}/T}$ with $\mu^R$ for right-movers $(p>0)$ and $\mu^L$ for left-movers ($p<0$), and we used that $\mathcal{A}(p) = \mathcal{B}p/2m^*T$. To linear order in $\Delta\mu=\mu^R-\mu^L$ one finds   
\begin{equation}\label{eq:NR-mu}
\dot{N}^R=-\frac{\Delta\mu}{\pi}\frac{L}{l_{\text{bs}}},\quad l^{-1}_{\text{bs}}=\frac{\mathcal{B}}{\sqrt{8\pi m^*T^3}}e^{-\varepsilon_F/T}.
\end{equation}
The exponential factor in the backscattering length $l_{\text{bs}}$ is due to the small statistical occupation probability of holes 
at the bottom of the band, while the prefactor reflects the fact that backscattering occurs in a diffusion process 
in momentum space. 

While the Fokker-Planck analysis holds in short wires, backscattering processes in long wires transform the unperturbed distribution of particles to a partially equilibrated form with finite boost velocity $u$. To identify the relevant length scale and to establish a relation between $\dot{N}^R$ and $u$ one can employ energy conservation. In a fully equilibrated wire with $\Delta\mu=0$ and $u=I/en$,
the backscattering in momentum space from the right to the left Fermi point does not cost energy but requires a net momentum $2p_F$ to be redistributed between right- and left-movers. This momentum is equally absorbed by particle-hole excitations at the Fermi level.  That is, creating and destroying particle-hole pairs with energy $v_Fp_F$ at the right and left Fermi point, respectively. As a result of the backscattering process, the energy balance for the right-movers thus consists of a loss of $\varepsilon_F$ due to removal of one particle from the Fermi level and a gain of $2\varepsilon_F$ due to the redistribution of momentum. Thus per each act of backscattering $\Delta N^R=-1$ the energy changes by $\Delta E^R=\varepsilon_F$, which gives a fixed relationship between the respective rates 
\begin{equation}\label{eq:ER-NR}
\dot{E}^R=-\varepsilon_F\dot{N}^R.
\end{equation}
If we now introduce the heat current $j_Q=j_E-\mu j$, as the energy current counted from the Fermi energy with $j$ the particle current, 
the energy flux gives the equation $j_Q=\dot{Q}^R$ where $\dot{Q}^R=\dot{E}^R-\mu\dot{N}^R$. The heat current is then readily calculated from the fully equilibrated electronic Fermi distribution function in the boosted frame of reference with $u=I/en$. Carrying out a Sommerfeld expansion to leading order in $T/\varepsilon_F$, one finds 
\begin{equation}\label{eq:jQ}
j_Q=\frac{\pi^2T^2}{6\varepsilon_F}nu,
\end{equation}
where at our level of accuracy we we do not distinguish between $\mu$ and $\varepsilon_F$. When combined with the preceding considerations on the energy balance this finally gives   
\begin{equation}\label{eq:NR-u}
\dot{N}^R=-\frac{\pi^2T^2}{12\varepsilon^2_F}nu.
\end{equation}
This equation together with Eq. \eqref{eq:NR-mu} fixes the relation between $\Delta \mu$ and $u$. In combination with Eqs. \eqref{eq:Landauer-G} and \eqref{eq:I-mu-u} this then determines the conductance of interacting electrons beyond the pure Luttinger liquid model 
\begin{equation}\label{eq:G}
\mathcal{G}=\mathcal{G}_0\left[1-\frac{\pi^2}{12}\frac{T^2}{\varepsilon^2_F}\frac{L}{L+l_\text{eq}}\right],
\end{equation}
where we introduced the relevant length scale for equilibration $l_{\text{eq}}=(\pi^2T^2/12\varepsilon^2_F)l_{\text{bs}}$. 
As alluded to in the introduction of this section, the interaction correction to the conductance of a Fermi-Luttinger liquid 
saturates to the value $\delta\mathcal{G}/\mathcal{G}_0\sim-(T/\varepsilon_F)^2$ in the long wire limit $L\gg l_{\text{eq}}$. For practical applications,  we should notice however that it is   hard to exceed the equilibration length with the experimentally available quantum wire micro-constrictions. In the currently accessible domain of parameters therefore $L<l_{\text{eq}}$, and the correction remains activated in temperature, $\delta\mathcal{G}/\mathcal{G}_0\sim-(T/\varepsilon_F)^2(L/l_{\text{eq}})\propto e^{-\varepsilon_F/T}$.

\subsection{Thermal response}\label{sec:thermal}

Let us next consider that the quantum wire is thermally biased by a temperature difference $\Delta T$ 
applied to the leads. A similar consideration of energy fluxes carried by left- and right-movers gives the Landauer type formula 
for the thermal conductance  
\begin{equation}\label{eq:Landauer-K}
\mathcal{K}_0\Delta T=j_Q-\dot{Q}^R.
\end{equation}
Here $\dot{Q}^R$ is the rate at which right-movers release their energy to left-movers via multi-particle thermalizing collisions. 

The rate of thermalization $\dot{Q}^R$ receives contributions from both, chirality conserving collisions and backscattering processes. In a parametrically wide range of lengths, $l_{\text{th}}<L<l_{\text{bs}}$, backscattering is weak and can be neglected. In this regime the electrical conductance remains unchanged by interactions. The thermal conductance, in contrast, is modified by thermalization due to the 
RRL quasiparticle decay processes discussed in Sec. \ref{sec:qp-decay}. More specifically, the correction due to these processes 
is found to be $\delta\mathcal{K}/\mathcal{K}_0\simeq -(T/\varepsilon_F)^2$ \cite{AL-K}. 

A more dramatic effect occurs in long wires due to backscattering collisions. To explore this limit, recall that the thermal conductance is defined as the proportionality coefficient between the heat flux current and temperature difference across the wire, computed at zero electrical current $j_Q=\mathcal{K}\Delta T|_{I=0}$. It then follows from Eq. \eqref{eq:I-mu-u} that to nullify the current the local difference in chemical potentials of left- and right-movers is locked to the drift velocity, $\Delta\mu/\pi=-nu$. Using Eqs. \eqref{eq:NR-mu} and \eqref{eq:ER-NR} the rate of thermalization in short wires $L\ll l_{\rm eq}$, can be brought to the form
\begin{equation}
\dot{Q}^R=-2\varepsilon_F\frac{L}{l_{\text{bs}}}nu, 
\end{equation}
and we notice again that the difference between the chemical potential $\mu$ and Fermi energy $\varepsilon_F$ is irrelevant for this discussion. Finally, expressing $nu$ via the heat current $j_Q$ from Eq. \eqref{eq:jQ} and inserting this into Eq. \eqref{eq:Landauer-K}, one finds the thermal conductance for arbitrary wire lengths \cite{Micklitz-Rech}
\begin{equation}
\mathcal{K}=\mathcal{K}_0\frac{l_{\text{eq}}}{L+l_{\text{eq}}}.
\end{equation}
A notable feature of this result is that the thermal conductance vanishes for long wires, $L\gg l_{\text{eq}}$. 
Nevertheless the thermal conductivity $\kappa=\mathcal{K} L$ remains finite, 
\begin{equation}
\label{kappa}
\kappa=\frac{\pi}{3}Tv_F\tau_{\text{eq}}.
\end{equation}
Clearly the large thermal conductivity is a direct manifestation of the exponentially slow relaxation processes in the one-dimensional system. The result Eq.~\eqref{kappa} also applies to the Luttinger liquid regime of strongly interacting fluids, upon replacing $v_F\to v$ the velocity of
bosonic excitations and a respective change of the equilibration time.

One should keep in mind that the above expression applies to the thermal conductivity defined at the lowest frequencies $\omega\ll \tau^{-1}_{\text{eq}}$ of measurement. However, the existence of hierarchical stages of relaxation processes
discussed in Sec. \ref{sec:rates}   opens a broad range of frequencies, $\tau^{-1}_{\text{eq}}\ll\omega\ll\tau^{-1}_{\text{qp}}$, where the thermal conductivity is governed by the gas of elementary quasiparticle excitations of the quantum liquid. Its value can be calculated from the collision integral of triple collisions Eq. \eqref{eq:St-eee}, focusing on RRL and LLR processes that enable energy exchange between excitations of different chirality. When solving the kinetic equation, one has to take care of projecting out all zero modes of the collision operator. These include in the present situation not only those related to conservation of particle number $(N)$, total momentum $(P)$ and energy $(E)$, but also conservation of the difference in the number of right- and left-movers $N^R-N^L$.
A solution method of the corresponding kinetic problem was presented in Ref. \cite{AL-K}. To evaluate $\kappa$ in this regime it is further sufficient to set $\tau^{-1}_{\text{eq}}\to0$ first, and only then take the limit $\omega\to0$. The calculation then gives  
\begin{equation}
\kappa\simeq T (T/\varepsilon_F)^2v_F\tau_{\text{qp}}
\end{equation}
modulo a numerical factor of order unity, in agreement with recent studies \cite{Matveev-Ristivojevic,Samanta} of the spinless model. 
Recalling that for weakly interacting spin-$1/2$  fermions $\tau^{-1}_{\text{qp}}\propto T$, we thus arrive at the temperature dependence 
$\kappa\propto T^2$ for the thermal conductivity of elementary excitations in the Fermi-Luttinger liquid. This drastically contrasts e.g. the thermal conductivity $\kappa\propto 1/(T\ln T)$ of a 2D Fermi gas \cite{Mishchenko}.

\subsection{Spin-charge energy partitioning}

The upshot of our discussion from the previous sections is that thermal effects strongly influence electrical transport in long wires. This is related to backscattering processes and the need to redistribute momentum among excitations. This physics becomes especially intricate in the Luttinger liquid regime as it relates to uneven energy partitioning between counterpropagating spin and charge modes \cite{Refael}. 
In this section we follow Ref. \cite{Matveev-Andreev-LLG} and repeat their arguments generalizing the above calculation of conductance to spin-charge coupled quantum fluids in the limit of arbitrary strength of interaction.  

We start out by noting that the total momentum of a Luttinger liquid is given by a sum of two contributions $P=p_FJ+P_b$ \cite{Haldane}. Here the first term represents the momentum of a filled Fermi surface with $J=N^R-N^L$ extra electrons at the right Fermi point, and the second term $P_b=P_\rho+P_\sigma$ captures the net momentum of (spin/charge) bosons. The momentum conservation combined with the particle number conservation, $N=N^R+N^L$, dictates that the rate of backscattering is related to relaxation of bosonic momentum 
\begin{equation}
\dot{N}^R=\dot{J}/2=-\dot{P}_b/2p_F.
\end{equation}
The momentum of bosons can be calculated from respective distribution in a boosted frame
\begin{equation}
P_b=-\sum_qq(qu)(\partial_\omega N^\rho_q+\partial_\omega N^\sigma_q)=\frac{\pi LT^2}{3}\left(\frac{1}{v^3_\rho}+\frac{1}{v^3_\sigma}\right)u.
\end{equation}
The relaxation of velocity $u$ of the boson gas towards drift velocity of the fully equilibrated wire $v_d=I/en$
\begin{equation}
\dot{u}=-\frac{u-v_d}{\tau_{\text{eq}}}
\end{equation}
is dominated by the backscattering in the spin-spin channel $\tau^{-1}_{\text{eq}}\propto \mathcal{B}_{\sigma\sigma}e^{-\varepsilon_\sigma/T}$. It is also important to note that the boost is the same for both spins and changes as spin-charge drag scattering occurs on a time scale $\tau_{\rho\sigma}$, that is much faster than the equilibration time. Next, one notices that the heat flux density determines the rate of energy exchange in backscattering processes as
\begin{equation}
j_Q=\dot{E}^R=\frac{\pi^2T}{3}\left(\frac{1}{v_\rho}+\frac{1}{v_\sigma}\right)u.
\end{equation}
Finally, in order to close the loop of formulas we need a relation between $\dot{E}^R$ and $\dot{N}^R$. Consideration of the energy partitioning between spin and charge modes propagating at different velocities suggests \cite{Matveev-Andreev-LLG,Refael} 
\begin{equation}
\dot{E}^R=-\frac{v_\rho v_\sigma(v^2_\rho+v^2_\sigma)}{(v^3_\rho+v^3_\sigma)}p_F\dot{N}^R. 
\end{equation}
Here the fraction results from the momentum proportional to $v^{-3}_{\rho/\sigma}$, transferred into the charge or spin branch, whereas the corresponding energies scale as $v^{-2}_{\rho/\sigma}$. Following then the calculation for the weak coupling limit, one can now combine the above relations to determine $\dot{N}^R$ to linear order in $v_d$. From this we then find the conductance with help of Eq. \eqref{eq:Landauer-G}. Focusing on long wires, $L\gg l_{\text{eq}}$, the interaction-induced correction can be cast in the form  
\begin{equation}\label{eq:deltaG}
\frac{\delta\mathcal{G}}{\mathcal{G}_0}=-\frac{\pi^2}{6}\frac{T^2}{v^2_\sigma p^2_F}F(v_\sigma/v_\rho), \ F=\frac{1+x^2}{(1+x)(1+x^3)}.
\end{equation}
This result is structurally identical to Eq. \eqref{eq:G} when taken in the same limit. Eq. \eqref{eq:deltaG} indicates that spin excitations give a dominant contribution to the interaction-induced correction to the conductance, as $F\to1$ for $v_\rho\gg v_\sigma$. Notice also that in this limit $\delta\mathcal{G}$ becomes independent of characteristics of the charge mode. For shorter wires, $L\ll l_{\text{eq}}$, the correction still shows activated temperature dependence $\delta\mathcal{G}\propto e^{-\varepsilon_\sigma/T}$. 
The activation energy $\varepsilon_\sigma\ll\varepsilon_F$ is set by spin excitations, and thus reduced by strong interactions compared to the weak coupling limit. 

\subsection{Momentum relaxing processes}

For completeness, we next consider extrinsic mechanisms of momentum relaxation that compete with equilibration processes and thus affect temperature dependence of resistance in quantum wires. Specifically, we here focus on electron-phonon scattering and the effect of smooth long-range inhomogeneities. The latter is also relevant to the hydrodynamic regime of transport in strongly correlated electron liquids \cite{Andreev}. We do not touch upon relaxation in linear Luttinger liquids with point-like impurities, which has e.g. been studied in Ref. \cite{Bagrets}. 

To calculate the phonon-induced resistivity we adopt the approach of  Sec. \ref{sec:el-ph}, used there to determine the relaxation time of hot electrons. Indeed, in the presence of equilibration processes the distribution function of electrons in the wire develops a finite drift velocity $u$, and right-movers become effectively hotter than left-movers. Technically, this becomes evident noting that the finite drift velocity enters the Fermi distribution as $pu$, which for states near the Fermi points $\pm p_F$ can be absorbed into a redefinition of distinct temperatures  
$T\pm \Delta T^{R/L}=T\pm T(u/v_F)$ for right- and left-movers, respectively. We can then calculate from Eq. \eqref{eq:St-ep} the rate of momentum loss by e.g. right-movers. Focusing on the phonon emission,
\begin{equation}
\dot{P}^R_{\text{ep}}=2\pi\sum_{pp'q}(p'-p)W[n_{p}(1-n_{p'})(1+N_q)-n_{p'}(1-n_p)N_q]. 
\end{equation} 
We find upon linearization in the drift velocity 
\begin{equation}
\dot{P}^R_{\text{ep}}=-\frac{2\pi u}{v_F}\sum_{pp'q}\frac{q_x}{T}W\epsilon_pf_{\epsilon_p}(1-f_{\epsilon_p})(1-f_{\epsilon_p-\varpi_q}-f_{\epsilon_p+\varpi_q}).
\end{equation} 
Here we employed momentum and energy conservations implicit in $W$, and introduced the following notations for energy variables $\epsilon_p=v_Fp$ and $\varpi_q=v_Fq_x$. The momentum integral over $p'$ is removed by the momentum conservation. The other momentum integration, $\sum_p\to\frac{L}{2\pi v_F}\int d\epsilon_p$, can be completed    
in the closed form with the help of the tabulated integral 
\begin{equation}
\int\limits^{+\infty}_{-\infty}\frac{xdx}{\cosh^2(x)}[\tanh(x+y)+\tanh(x-y)]=\frac{2y^2}{\sinh^2(y)}.
\end{equation} 
Imposing a force balance condition on segments of the wire shorter than $\xi$, we define the resistivity 
$\rho_{\text{ep}}=-\dot{P}^R_{\text{ep}}/(e^2n^2uL)$. The latter is then found to be of the form 
\begin{equation}
\rho_{\text{ep}}=\frac{1}{e^2n^2v^2_F}\sum_q\frac{q_x}{4T}W(q)\frac{(v_Fq_x)^2}{\sinh^2(v_Fq_x/2T)}.
\end{equation} 
Finally, we separate in the phonon dispersion momentum components along and perpendicular to the wire, $\omega_{q}=s\sqrt{q^2_x+q^2_\perp}$, and notice that energy conservation fixes $q_x\approx (s/v_F)q_\perp$. That is, because of the large difference of phonon and electron velocities, phonons are emitted predominantly in transverse direction with respect to the electron momentum. The final integral over phonon phase space is  separately done for piezoelectric and deformation potentials contributing to the scattering rate $W$. The result is thus a sum of the following two contributions,
$\rho_{\text{ep}}=\rho_{\Lambda}+\rho_{D}$, where 
\begin{subequations}\label{eq:-rho-ep}
\begin{align}
\rho_{\Lambda}=\frac{3\zeta(3)}{8\pi^3e^2}\left(\frac{T}{v_F}\right)^2\left(\frac{T}{s}\right)\frac{\langle\Lambda^2\rangle}{n^2\varrho sv^2_F},\\
\rho_D=\frac{15\zeta(5)}{\pi^2e^2}\left(\frac{T}{v_F}\right)^4\left(\frac{T}{s}\right)\frac{D^2}{n^2\varrho s^3}.
\end{align}
\end{subequations}
Here we approximated the piezoelectric coupling function by its average over angles of the emitted phonons in transverse direction, namely
$\langle\Lambda^2\rangle=\int^{2\pi}_{0}\Lambda^2(0,q_\perp\cos\phi,q_\perp\sin\phi)d\phi$. 

In addition to electron-phonon scattering, two-particle electron collisions may also provide a channel for momentum relaxation  
when translational invariance of the system is broken. This happens e.g. when screening of the interaction is inhomogeneous. The corresponding relaxation process is captured by the usual collision integral
\begin{align}\label{eq:PR-ee}
&\dot{P}^R_{\text{ee}}=2\pi\sum_{\stackrel{pp'>0}{kk'<0}}(p'+k'-p-k)W\nonumber 
\\ &\left[n_pn_k(1-n_{p'})(1-n_{k'})-n_{p'}n_{k'}(1-n_p)(1-n_k)\right],
\end{align}
where crucially the scattering rate $W$ does not preserve momentum. Notice also that momenta in Eq.~\eqref{eq:PR-ee} are restricted 
to include one particle-hole excitation in each branch. To accommodate inhomogeneous screening in the simplest model, 
one can assume an interaction potential with separable kernel $V(x,x')\to V(x-x')\Upsilon((x+x')/\xi)$.
Here the first term represents the usual screened Coulomb interaction, and the nonuniformity of the system is encoded in the dimensionless function $\Upsilon(x)$. The latter has a spatial scale with correlation radius $\xi$, which we assume to be large compared to both, the Fermi wavelength and range of interaction $V(x-x')$. The matrix element of interaction $V_{pk,p'k'}=\langle k'p'|V(x,x')|kp\rangle$ entering the scattering rate $W$ can be calculate in the basis of plane waves $|pk\rangle=\frac{1}{\sqrt{2}}(e^{ipx}e^{ikx'}-e^{ipx'}e^{ikx})$. A more accurate analysis based on quasiclassical wave functions leads to qualitatively the same result. In complete analogy to electron-phonon scattering, one finds upon expansion of Eq.~\eqref{eq:PR-ee} to linear order in the boost velocity the resistivity 
\begin{align}
\rho_{\text{ee}}=\frac{(T^3/\varepsilon_Fv^3_F)}{2\pi^2e^2n}\left(\frac{V_0-V_{2p_F}}{v_F}\right)^2\int\limits^{+\infty}_{-\infty}\frac{(\omega/2T)^4d\omega}{\sinh^2(\omega/2T)}\nonumber \\ \times
\frac{1}{\Delta x}\int\limits^{x+\Delta x}_{x} dxdx'\Upsilon(x/\xi)\Upsilon(x'/\xi)e^{-2i\omega(x-x')/v_F}.
\end{align}
To arrive at the above expression we split the energy conserving delta function in the scattering rate into two, 
$\delta(\varepsilon_p+\varepsilon_k-\varepsilon_{p'}-\varepsilon_{k'})\to\int d\omega\delta(\varepsilon_p-\varepsilon_{p'}-\omega)\delta(\varepsilon_k-\varepsilon_{k'}+\omega)$, and made use of the following identities applicable for equilibrium Fermi functions,
\begin{align}
f_\varepsilon(1-f_{\varepsilon\pm\omega})=\frac{f_{\varepsilon}-f_{\varepsilon\pm\omega}}{1-e^{\mp\omega/T}}, \int\limits^{+\infty}_{-\infty} d\varepsilon (f_\varepsilon-f_{\varepsilon\pm\omega})=\pm\omega.
\end{align}
These steps enable us to complete three  out of the four momentum integrations in Eq. \eqref{eq:PR-ee}. The final result for $\rho_{\text{ee}}$ depends on the relation between temperature $T$ and the characteristic energy scale of the inhomogeneity,
$E_\xi=v_F/\xi$. In the limit of high temperatures, $T\gg E_\xi$, the exponential under the integral is a rapidly oscillating function. 
One can explore this fact to integrate by parts, using that $(\omega/v_F)^2e^{-2i\omega(x-x')/v_F}=\frac{1}{4}\partial^2_{xx'}e^{-2i\omega(x-x')/v_F}$, and then notice that the remaining frequency integral gives a delta function $\delta(x-x')$. 
As a result one arrives at
\begin{equation}\label{eq:rho-ee}
\rho_{\text{ee}}=\frac{1}{32\pi e^2}\left(\frac{V_0-V_{2p_F}}{v_F}\right)^2\frac{T}{n\varepsilon_F}\left(\partial_x\Upsilon(x)\right)^2,
\end{equation}
where the remaining position integral was expanded to first order in $\Delta x$. In the opposite limit of low temperatures, $T\ll E_\xi$,
one finds the weaker dependence $\rho_{\text{ee}}\propto T^4$. 

The above calculation for the weakly interacting limit can be generalized to the Luttinger liquid regime. The inhomogeneity induced resistivity is then given by a sum of two contributions, $\rho_{\text{ee}}=\rho_\rho+\rho_\sigma$, as both charge and spin modes dissipate energy throughout the wire. At $T\gg E_\xi$ both contributions remain linear in temperature $\rho_{\rho/\sigma}\propto T$ \cite{Rech}. 

A crucial point to re-emphasize is that both extrinsic mechanisms leading to resistivities $\rho_{\text{ep}}$ and $\rho_{\text{ee}}$ 
are made possible only by electron backscattering which establishes a finite boost $u$ in the distribution function 
of partially equilibrated electrons. In order to find the complete wire resistance one needs to modify the 
energy balance condition Eq. \eqref{eq:ER-NR} to account for both momentum relaxing collisions
\begin{equation}
\dot{E}^R=-\mu\dot{N}^R+v_F\dot{P}^R_{\text{mr}},\quad \dot{P}^R_{\text{mr}}=\dot{P}^R_{\text{ep}}+\dot{P}^R_{\text{ee}}.
\end{equation}
Since both $\dot{P}^R_{\text{ep}}$ and $\dot{P}^R_{\text{ee}}$ are proportional to $u$ this condition mixes $\Delta \mu$ and $u$ in the energy balance. It is then useful to introduce the length scale  $l_{\text{mr}}$ for momentum relaxation
\begin{equation}
\dot{P}^R_{\text{mr}}=-unp_F\frac{L}{l_{\text{mr}}},\quad l^{-1}_{\text{mr}}=\frac{e^2}{\pi}(\rho_{\text{ep}}+\rho_{\text{ee}}).
\end{equation}
Recalling Eq. \eqref{eq:rho-ee} and its weak coordinate dependence, $\rho_{\text{ee}}$ should here be understood as averaged 
over the wire. Repeating then the steps from Sec. \ref{sec:electrical}, that is, excluding $\Delta \mu$ with help of energy balance equations and backscattering rates, and using Eq. \eqref{eq:I-mu-u} to express $u$  in terms of the current through the wire $I$, one finally finds the wire resistance $R=V/I$ from Eq. \eqref{eq:Landauer-G}, 
 \begin{equation}
 R=\frac{\pi}{e^2}\left[1+\frac{r_{\text{bs}}(r_0+r_{\text{mr}})}{(r_{\text{bs}}+r_0+r_{\text{mr}})}\right].
 \end{equation}
Here we introduced the short-hand notations $r_0=\pi^2T^2/12\varepsilon^2_F$, $r_{\text{bs}}=L/l_{\text{bs}}$, and $r_{\text{mr}}=L/l_{\text{mr}}$. For long wires $r_0\ll \{r_{\text{mr}},r_{\text{bs}}\}$, and one finds a resistance scaling linearly with the wire length, 
$R=\rho L$, with resistivity given by    
 \begin{equation}
 \label{rho_total}
 \rho=\frac{\pi}{e^2}\frac{1}{l_{\text{bs}}+l_{\text{mr}}}.
 \end{equation}
This formula is in stark contrast to Matthiessen's rule, stating that resistivity $\rho=\rho_1+\rho_2+\ldots$ comprises of a sum of partial resistivities $\rho_i\propto l^{-1}_{i}$ originating from different scattering channels, each being inversely proportional to the respective scattering length $l_{i}$. The reason for this significant difference is that Matthiessen's rule misses both quantum and classical correlational effects. In the present context, the channels of chirality changing electron collisions and momentum relaxing electron collisions are strongly intertwined and Matthiessen's rule does not apply. That is, resistivities for competing relaxation channels 
in Eq.~\eqref{rho_total} add up in parallel rather than in series.

\subsection{Plasmon thermal transport}

To further elucidate an important difference between charge and thermal transport in quantum 1D liquids we consider thermal currents carried by bosonic excitations. In this section we use the framework developed in Refs. \cite{Fazio,Gutman-Gefen-Mirlin} and our condense discussion only to the spinless case of  inhomogeneous Luttinger liquids. 

We start out from the Landauer formula for the heat flux of plasmons  
\begin{equation}
j_E=\int^{\infty}_{0}\frac{d\omega}{2\pi}\omega|t_\omega|^2\left[N^R_\omega-N^L_\omega\right]
\end{equation} 
where all the effects of interactions in the wire are captured by the transmission coefficient of plasmons $|t_\omega|^2$. 
The notation $N^{R/L}_\omega$ refers to bosonic distributions for right/left moving excitations. For a small temperature difference $\Delta T$ maintained between the reservoirs we obtain thermal conductance in the form 
\begin{equation}\label{eq:K-plasmons}
\mathcal{K}=\frac{j_E}{\Delta T}=\int^{\infty}_{0}\frac{d\omega}{2\pi}\frac{(\omega/2T)^2}{\sinh^2(\omega/2T)}|t_\omega|^2
\end{equation}
In the noninteracting limit $|t_\omega|^2\to1$ this expression gives the thermal conductance quantum $\mathcal{K}_0$ of a perfect channel. Another simple limit to check is for a step-like model, with interaction parameter $g<1$ in the wire and $g=1$ in the leads. The transmission coefficient of plasmons in this case is that of the Fabri-P\'erot interferometer
\begin{equation}\label{eq:t-Fresnel}
|t_\omega|^2=\frac{1}{1+\left(\frac{g^2-1}{2g}\right)^2\sin^2(2\pi\omega/\omega_L)}.
\end{equation}

At high temperatures $T\gg\omega_L=v/L$ the transmission coefficient in Eq.~\eqref{eq:K-plasmons} strongly oscillates. It can then be replaced by its average value $\langle|t_\omega|^2\rangle=\frac{1}{\omega_L}\int^{\omega_L}_{0}|t_\omega|^2d\omega=\frac{2g}{g^2+1}$, which leads to the thermal conductance $\mathcal{K}=2g\mathcal{K}_0/(g^2+1)$ quoted at the beginning of this section.

With these elementary examples, we next discuss interaction correction to the thermal conductance  in an inhomogeneous wire. We here focus on the model formulated in Ref. \cite{Fazio} and restrict analysis to the lowest temperatures. To this end, we consider the wave equation for plasmons in the basis of scattering states 
\begin{equation}
-\hat{h}\phi^{R/L}_{k}(x)=\omega^{2}_{k}\phi^{R/L}_{k}(x)\,,
\end{equation}
where $L/R$ labels left- and right-movers, $\omega_{k}$ is the energy of the given mode, and
\begin{equation}
\hat{h}=\sqrt{n(x)}\partial_{x}\left(V(x)/m+\pi^2n(x)/m^2\right)\partial_{x}\sqrt{n(x)}\,.
\end{equation}
is the Hamiltonian operator of the inhomogeneous wire with coordinate dependent interaction potential. The scattering states of left-movers take the asymptotic form 
\begin{equation}
\phi^L_{k}(x)=\left\{
\begin{array}{lc}
e^{ikx}+r_{\omega_k}e^{-ikx} & x\to-\infty \\
t_{\omega_k}e^{ikx} & x\to+\infty
\end{array}
\right.
\end{equation}
and similar for the right-movers. To establish a link to the model employed in the previous sub-section we further treat the electron density $n(x)$ as constant and only allow space dependence of the interaction in the model of a separable kernel. In this case the scattering problem simplifies to
\begin{equation}
(v^2\partial^{2}_{x}+\omega^{2}_{k})\phi^{L/R}_{k}(x)=-\frac{v}{\pi}\partial_{x}(V(x)\partial_{x}\phi^{L/R}_{k}(x)).
\end{equation}
We next treat this problem perturbatively in interaction and calculate leading order correction to the transmission coefficient. For this purpose 
we first construct the Green's function of the wave equation 
\begin{equation}
(v^{2}\partial^{2}_{x}+\omega^{2}_{k})G(x,y)=\delta(x-y)\,,
\end{equation}
which reads
\begin{equation}
G(x,y)=\frac{1}{2ikv^2}e^{ik|x-y|}.
\end{equation}
We then recall that for weak interactions in the model under consideration, $V\to (V_0-V_{2p_F})\Upsilon(x)$, and that for our purposes we can replace $v\to v_F$. As a result, the solution for the scattering problem can be cast in the form
\begin{align}
&\phi^L_{k}(x)=e^{ikx}+\delta\phi^L_{k}(x), \nonumber \\ 
&\delta\phi^L_{k}(x)=-\frac{V_0-V_{2p_F}}{2\pi ikv_F}\int dy\,
e^{ik|x-y|}\partial_{y}(\Upsilon(y)\partial_{y}e^{iky})\,.
\end{align}
Since for the reflected wave $\delta\phi^L_{k}(x)=r_{\omega_k}e^{-ikx}$ we take the asymptotic of
$\delta \phi_{k,L}(x)$ at $x\to-\infty$ and determine the reflection
coefficient to be
\begin{equation}
r_{\omega_k}=\frac{i\omega_k}{2\pi v^2_F}(V_0-V_{2p_F})\int dy\,
\Upsilon(y)e^{2iky}\,.
\end{equation}
Noting that $|t_\omega|^2=1-|r_\omega|^2$, it is evident from Eq. \eqref{eq:K-plasmons} that corrections to the thermal conductance $\delta\mathcal{K}$ are due to plasmon backscattering. Inserting now $|r_\omega|^2$ under the integral we get
\begin{equation}
\frac{\delta\mathcal{K}}{\mathcal{K}_0}=-\frac{c^2_\Upsilon}{5}\left(\frac{V_0-V_{2p_F}}{v_F}\right)^2
\left(\frac{T}{E_\xi}\right)^2
\end{equation}
where $c_\Upsilon$ is dimensional model specific number. This result applies to lowest temperatures, $T\ll E_\xi$, where $e^{2i\omega(y_1-y_2)/v_F}\sim 1$ since $\omega\sim T$ and $y-y'\sim \xi$. The same result can be derived in fermionic language from the kinetic equation, by calculating $\dot{Q}^R$ based on two-particle momentum relaxing scattering enabled by the inhomogeneity. As compared to a similar calculation of resistivity $\rho_{\text{ee}}$, here the correction to the thermal transport coefficient emerges without electron backscattering. The connection to a step-function model is clear as well. Indeed, from Eq. \eqref{eq:t-Fresnel} one can expand the transmission in the limit $\omega/\omega_L\ll1$ and retain the leading order interaction renormalization, $g\sim 1-V_0/v_F$, which consistently yields a correction $\delta\mathcal{K}\propto T^3$. 

The complete picture of plasmon thermal transport in inhomogeneous 1D systems is rather complicated. Disorder tends to localize plasmons and their transmission then may get exponentially suppressed $|t_\omega|^2\sim e^{-L/\zeta_\omega}$ on the length scale $\zeta_\omega$ for plasmon localization. This process competes, however, with inelastic plasmon-plasmon scattering. For instance, two-boson $\rho\rho\to\rho\rho$ scattering, which as we discussed above in Sec. \ref{sec:spin-charge}, kinematically produce low energy excitations. Since the decay of plasmons diverges inversely proportional to the plasmon frequency, $\zeta_\omega\propto1/\omega^2$, such low-energy bosons avoid localization and propagate ballistically throughout the wire with almost perfect transmission. As a result, it should be expected that the thermal conductance depends in a non-trivial fashion on the system size $L$ (see Ref. \cite{Bard-Protopopov} for recent progress on this issue). A generalization of these effects to systems with spin degrees of freedom
is still an open problem.

\begin{figure}
\begin{center}
\includegraphics[width=.5\textwidth]{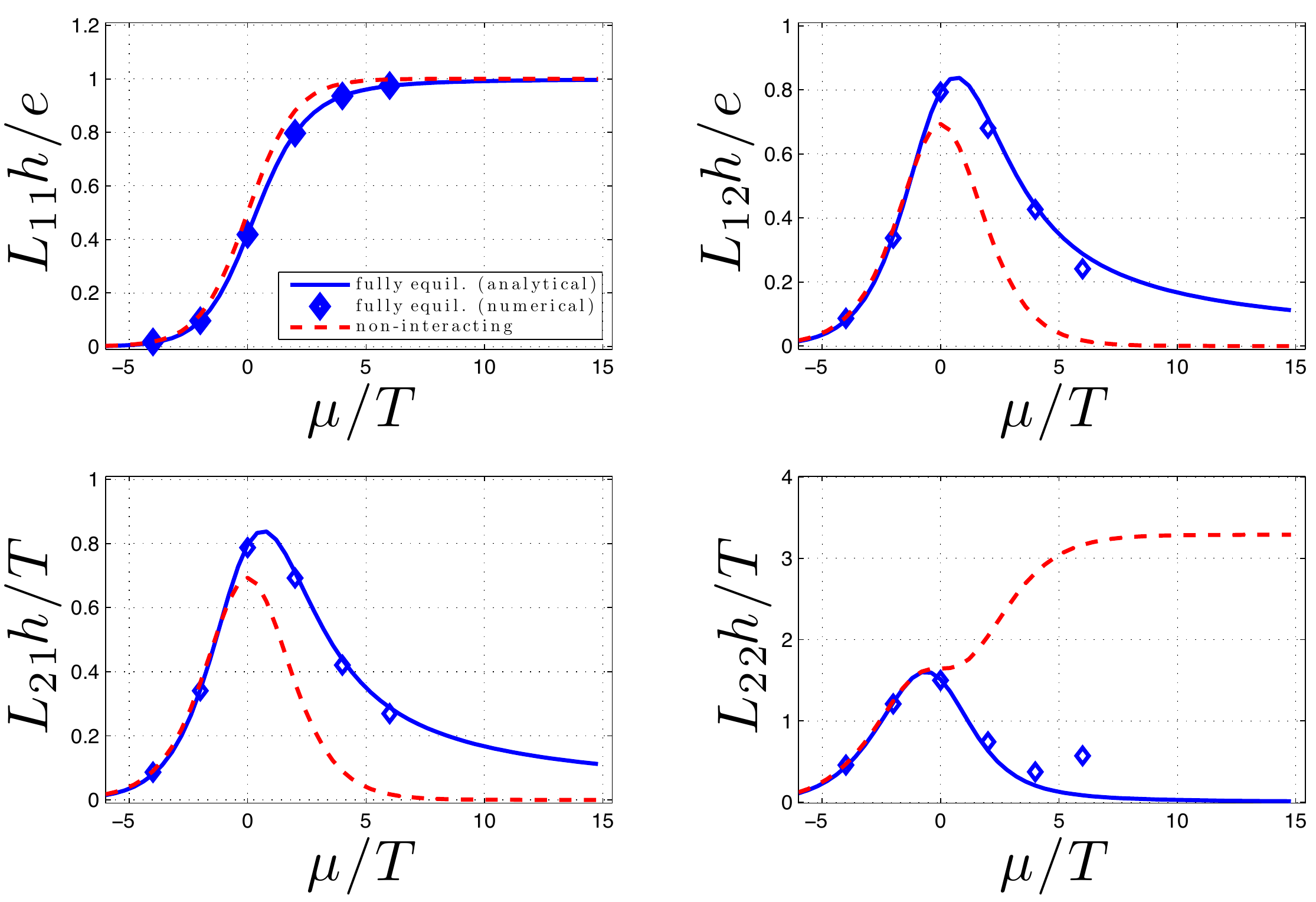}
\caption{Linear response transport coefficients at the QCP of the first plateau transition of fully equilibrated (solid line) and noninteracting electrons (dashed line). Numerical results (symbols) were obtained from the solution of the kinetic equation (see Ref.~\cite{Roch} for details). 
In all calculations we assumed the limit of long wires, $L\gg l_{\text{eq}}$, where $l_{\text{eq}}$ is the equilibration length at the QCP.}
\label{fig:L}
\end{center}
\end{figure}

\subsection{Thermoelectric matrix at QCP}

We next extend the analysis of the preceding sections to determine the full matrix of linear transport coefficients, 
\begin{align} 
\begin{pmatrix} 
j_c \\
j_Q
\end{pmatrix}
 = \hat{L}
 \begin{pmatrix}
eV\\
\Delta T
  \end{pmatrix}
\end{align} 
relating electric current ($j_c$) and heat current ($j_Q$) to voltage and temperature drops. We remind that components of this matrix define the thermopower $\mathcal{S}={ L_{12}\over eL_{11}}$, thermal conductance $\mathcal{K}=L_{22} - {L_{12}L_{21}\over L_{11}} $, Peltier coefficient $\Pi={L_{21}\over L_{11}}$, and electric conductance $\mathcal{G}=eL_{11}$, with the Onsager relation $\Pi=\mathcal{S}T/e$. 

The ideas that went into the calculation of the conductance of fully equilibrated quantum wires, viz. the  limit $L\gg l_{\text{eq}}$ of Eq. \eqref{eq:G}, relay exclusively on considerations of conservation laws. Thus one expects the conclusions of such analysis to be fairly generic. Indeed, in the spinless case, microscopic details of the interaction were not important in establishing energy partitioning between left- and right-movers in a backscattering process as they only went into the definition of the equilibration length. As a result, the interaction-induced correction to the conductance took a universal form. An interesting application of these ideas is to investigate the impact of equilibration processes on transport coefficients near the transition from one conductance plateau to the next \cite{Roch}. The latter is an example of a quantum phase transition. Importantly, in the vicinity of each plateau transition the equilibration length crosses from exponential to power-law behavior, thus making an experimental verification of equilibration effects on the conductance more promising. 
From a dimensional analysis  we estimate that for generic nonintegrable short-range interaction near the quantum critical point (QCP) $l^{-1}_{\text{eq}}\propto T^{13/2}$. 

Following the strategy outlined in Sec. \eqref{sec:electrical} the transport matrix is found to be of the form
\begin{align} 
\begin{pmatrix} 
j_c \\
j_Q
\end{pmatrix}
 = \frac{\mathrm{g}}{2\pi\hbar}
 \begin{pmatrix} 
e& e \beta \\
T \beta  & T \beta^2
\end{pmatrix}
 \begin{pmatrix}
eV\\
\Delta T
  \end{pmatrix}.
\end{align} 
Here we restored $\hbar$, and introduced the dimensionless conductance 
\begin{equation}\label{eq:G-QCP}
\mathrm{g}(z)={\alpha_1^2(z)-\alpha_2(z)\alpha_0(z)
\over 
2\alpha_1(z)\beta(z)-\alpha_2(z)-\alpha_0(z)\beta^2(z)}
\end{equation}
with $z=\mu/T$, where the set of functions 
\begin{equation}
\alpha_n(z) = \langle \chi^n \rangle_{z}, \quad \beta(z) =  { \langle \chi \sqrt{1 + \chi/ z} \rangle_{z}
\over
\langle \sqrt{1 +  \chi/ z }\rangle_{z}},
\end{equation}
is defined by the thermal average 
\begin{align}
\langle ... \rangle_{z}=  - \int_{-{z}}^{\infty} d\chi (...) {d f_\chi \over d\chi},\quad f_\chi=\frac{1}{e^\chi+1}.
\end{align}
The components of the thermoelectric matrix $L_{ij}$ for a fully equilibrated wire at the QCP are shown in Fig. \eqref{fig:L} in comparison to the noninteracting limit. It can be readily checked that away from the QCP, $z=\mu/T\gg1$, one finds to leading order in $1/z$ $\alpha_0=1$, $\alpha_1=0$, $\alpha_2=\pi^2/3$, and $\beta=\pi^2/6z$. This reproduces the saturated form of the conductance,
 $\mathrm{g}=1/(1+\pi^2/12z^2)$. In contrast, one finds at the QCP $z=\mu/T\to0$, $\alpha_0=1/2$, $\alpha_1=\ln 2$, $\alpha_2=\pi^2/6$, and $\beta=-\frac{3\zeta(3/2)}{2\sqrt{2}\zeta(1/2)}$. This implies that $\mathcal{G}_{\text{QCP}}=\mathrm{g}\frac{e^2}{2\pi\hbar}$ with $\mathrm{g}\approx0.42$, which is about $16\%$ below the result of the noninteracting limit. 
 
 These results can be readily generalized to the QCP of the $N^{\text{th}}$ plateau transition, where chemical potentials $\mu_i\gg T$ ($i=1,\ldots, N-1$) while $\mu_N\sim T$. 
 We can then determine that at the QCP $z_N=\mu_N/T\to0$, asymptotically $\mathcal{G}_{\text{QCP}}\approx\frac{e^2}{2\pi\hbar}\left[N-\frac{1}{2}-\frac{2\ln^22}{\pi^2(N-1)}\right]$, for $N>1$. We notice that the conductance of quantum wires was carefully measured in Ref. \cite{Cronenwett} for different temperatures. For quantum critical points of the first and second plateau transition, that correspond to those points where curves for different temperatures intersect, a reduced value of the conductance as compared to the value of noninteracting electrons was observed. This is in qualitative agreement with the presented here picture of interaction effects.   

\subsection{Spin-orbit effects and magnetoconductance}

Recent significant progress in fabrication of clean gated semiconducting InAs and InSb quantum wires with strong Rashba-type spin-orbit coupling \cite{Kammhuber,Gazibegovic} opens new avenues for exploring the consequences of the interplay in spin physics and electron interactions on the nanoscale 1D transport. The simplest model that describes these systems includes Rashba and Zeeman terms in the Hamiltonian. The former lifts spin degeneracy, shifting electron bands in momentum space according to the spin polarization, and the latter further shifts bands in energy, partially gapping the spectrum. The resulting single particle energy bands are  
\begin{equation}\label{eq:rashba-bands}
\varepsilon^\pm_p=\frac{p^2}{2m}\pm\sqrt{B^2_z+(\alpha_{\text{so}}p)^2}
\end{equation}   
where $\alpha_{\text{so}}$ describes the strength of Rashba coupling, and $B_z$ defines the energy scale of Zeeman splitting linear in magnetic field. From the stand point of transport properties, the most interesting configuration is when chemical potential is within the Zeeman gap $-B_z<\mu<B_z$. Then only the lowest helicity sub-band is occupied and the model can be bosonized. Linear Luttinger liquid theory predicts quantization of the conductance for such Rashba wires, adiabatically connected to noninteracting leads
at the value $e^2/h$, irrespective of the interaction strength in the wire \cite{Meng}. The reasoning is exactly the same as for conventional quantum wires discussed earlier. Taking into account curvature of the spectrum around the Fermi points within an extended model, the
peculiar nonlinear energy dispersion relation Eq. \eqref{eq:rashba-bands} allows for both two-particle and three-particle inelastic \textit{e-e} scattering processes. The impact of the latter on interaction-induced corrections to the quantized conductance  was investigated in the recent papers \cite{Schmidt-RashbaWire,Nagaev} for short quantum wires.          

To quantify these effects one needs to single out chirality changing processes. That is, scattering processes that do not conserve the number of right/left-movers before and after the collision. The leading kinematically allowed two-particle process involves a set of states within the same helicity band accompanied by an interband transition. Momentum and energy conservations dictate that the bottleneck for this process is the existence of an unoccupied state near $p\sim 0$. At finite chemical potential lying within the Zeeman gap this costs a Boltzmann factor $\propto e^{-B_z/T}e^{-|\mu|/T}$. Being a two-particle process, Golden rule dictates the correction to the conductance scaling in the 
interaction parameter $\propto V^2$. As usual, thermal broadening allows momentum transfer in the collision of order $\sim T/v_F$. However, for the dispersion relation with the square root Eq. \eqref{eq:rashba-bands}, we expect contribution with an enhancement $\propto 1/\sqrt{T}$ of the scattering processes. Combining these observations with dimensional reasoning, suggests that the quantum correction to conductance scales as
$\delta\mathcal{G}/\mathcal{G}_0\sim -L/l_\text{bs}$ with the backscattering length
\begin{equation}\label{eq:dG-ee-so}
l^{-1}_{\text{bs}}\sim
\frac{B_z}{\alpha_{\text{so}}}\left(\frac{V}{\alpha_{\text{so}}}\right)^2\sqrt{\frac{T}{B_z}}e^{-B_z/T}e^{-|\mu|/T}.
\end{equation}    
This estimate is supported by a detailed calculation \cite{Schmidt-RashbaWire}. In a similar spirit, one may estimate the contribution from three-particle processes. In contrast to the two particle collisions, the latter only involve states from the same helicity band. At the perturbative Golden rule level the correction to the conductance comes with a factor $\propto V^4$, a phase factor $\propto T^3$, and a thermal factor $\propto e^{-B_z/T}e^{|\mu|/T}$. As a result we expect again $\delta\mathcal{G}/\mathcal{G}_0\sim -L/l_\text{bs}$, 
now with a backscattering length
\begin{equation}\label{eq:dG-eee-so}
l^{-1}_{\text{bs}}\sim \frac{B_z}{\alpha_{\text{so}}}\left(\frac{V}{\alpha_{\text{so}}}\right)^4\left(\frac{T}{B_z}\right)^3e^{-B_z/T}e^{|\mu|/T}.
\end{equation}   
In the above estimate we took advantage of the fact that spin-orbit coupling lifts integrability constraints, and  point-like interaction  already lead to a finite relaxation rate. That is, there are no more subtle cancellations which could bring additional $T$-dependent prefactors.

For wires e.g. with $L\sim 2$ $\mu$m, $\alpha_{\text{so}}\sim 1$ eV\AA, $B_z\sim 1.5 H $ meV/T and $V/\alpha_{\text{so}}\sim1$,
these corrections become experimentally noticeable for magnetic fields $H\sim 5$ mT and temperatures $T\sim 50$ mK such that $B_z/T\sim 2$ where $\delta\mathcal{G}/\mathcal{G}_0\sim10^{-2}$. Furthermore, one can expect that triple electron processes may dominate provided that $0<|\mu|\lesssim B_z$. We argue that in this regime, and for long enough equilibrated wires, $L\gg l_{\text{bs}}$, the interaction correction to conductance $\delta\mathcal{G}$ saturates. 
 
The calculation of the interaction induced correction to the conductance of fully equilibrated Rashba wires is exactly analogous to the previous example of transport near a QCP. We only need to accommodate the different dispersion relation Eq. \eqref{eq:rashba-bands}. 
The conductance is thus still given by Eq. \eqref{eq:G-QCP} but now with the modified thermal factors   
\begin{equation}
\alpha_n(z) = \langle\langle \chi^n \rangle\rangle, \quad \beta(z) =   \langle\langle\chi p(\chi) \rangle\rangle/\langle\langle p(\chi)\rangle\rangle,
\end{equation}
and a redefined average 
\begin{align}
\langle\langle ... \rangle\rangle=  - \left(\int_{-{z_1}}^{\infty}-\int^{-z_1}_{-z_0} \right)d\chi (...) {d f_\chi \over d\chi}.
\end{align}
Here $z_1=(\mu+m\alpha^2_{\text{s0}}/2+B^2_z/2m\alpha^2_{\text{so}})/T$, $z_0=(\mu+B_z)/T$ and $p(\chi)$ is the dimensionless momentum corresponding to the dispersion in Eq. \eqref{eq:rashba-bands}. In the limit $|\mu|\lesssim\{T,B_z\}\ll m\alpha^2_{\text{so}}$,  
analytical progress is possible by expanding $\alpha_{n}$ and $\beta$ in  corrections algebraically small in $\{T,B_z\}/ m\alpha^2_{\text{so}}$, and neglecting all other exponential small terms. As a  result, we find the conductance for the fully equilibrated Rashba wire 
\begin{equation}
\mathcal{G}=\mathcal{G}_0\left[1-\frac{\pi^2}{12}\left(\frac{T}{m\alpha^2_{\text{so}}}\right)^2+
\frac{5\pi^2}{42}\frac{T^2B^2_z}{(m\alpha^2_{\text{so}})^4}\right],
\end{equation} 
where we also retained the leading order field-dependent correction. This result is analogous to Eq. \eqref{eq:G} taken in the limit $L\gg l_{\text{eq}}$, but perhaps surprisingly it leads to a negative magnetoresistance. 


\section{Final remarks and perspective}\label{sec:summary}

In this work we in part reviewed recent progress on theory of kinetic processes in Fermi-Luttinger liquids and presented new results. We provided a comprehensive discussion of quasiparticle relaxation mechanisms in single channel quantum wires, and unveiled their impact on the thermoelectric and magnetotransport properties. While the emphasis of this study was on electron liquids in quantum wires, there are other interesting variants of Luttinger liquids, left aside in this paper, where similar physics can be explored. To provide a broader perspective we highlight a few interesting examples.   

We saw that the temperature and energy dependence of quasiparticle relaxation times is extremely sensitive to details of the spectrum nonlinearities. This aspect of the problem becomes extremely intricate in chiral Luttinger liquids of the integer quantum Hall effect. The interplay of interactions and confinement leads to either spin- or charge-dominated mechanisms of edge reconstruction. As a result, relaxation rates of hot electrons injected into edge channels are significantly altered in different reconstruction scenarios \cite{FelixVonOppen}. Similar complications exist in the regime of the fractional quantum Hall effect, where the interplay of counterpropagating modes of reconstructed edges has dramatic consequences on relaxation mechanism, and ultimately the temperature dependence of the electric and thermal conductances \cite{Kane-Fisher-Polchinski,Protopopov-Gefen-Mirlin}. 

The quantum spin Hall effect gives rise to a helical version of Luttinger liquids \cite{Bernevig}. In these systems potential scatterers alone cannot prevent electrons from ballistic propagation along the edges. That is, backscattering is not permissible since counterpropagating states of the same energy form a Kramers doublet if time-reversal symmetry is preserved. Breaking, however, the latter by e.g. magnetic impurities, or breaking axial spin symmetry in the presence of strong spin-orbit effects, opens the possibility for various inelastic and spin-flip scattering processes, thus enabling quasiparticle relaxation and ultimately affecting the conductance \cite{Maciejko,Rachel,Lezmy,Yudson,Kainaris,Foster}.  

One could add to this list another member, namely the so-called spiral Luttinger liquids \cite{Braunecker}. This peculiar state may form
in quantum wires where a spontaneous ordering of nuclear spins at low temperatures produces an effective Rashba spin-orbit coupling, leading to a strongly nonlinear single-particle spectrum. As a consequence, inelastic scattering processes become possible, which should result in interaction-induced corrections to transport properties in these systems.

We thus expect this field to continue evolving in various fruitful directions. In particular, a comprehensive understanding of kinetic processes and time-scales for relaxation presented in this work provides a necessary ingredient in bridging to a hydrodynamic description of strongly correlated electron liquids. 

\subsection*{Acknowledgements} 

We acknowledge collaborations with A. Andreev, L. Glazman, T. Karzig, K. A. Matveev, F. von Oppen, J. Rech, M. T. Rieder, A. Rosch, and Z. Ristivojavic. We would like to thank Maxim Khodas for useful discussions, for reading the manuscript prior to submission and for providing comments. This work was supported by the U. S. Department of Energy (DOE), Office of Science, Basic Energy Sciences (BES) 
Program for Materials and Chemistry Research in Quantum Information Science under Award No. DE-SC0020313.
T.~M.~acknowledges financial support by Brazilian agencies CNPq and FAPERJ.


\appendix

\section{Notes on Bosonization}

This section is prepared as a supplementary material to the main text of the paper. Here we concentrate on the derivation of an effective Hamiltonian for nonlinear Luttinger liquids and construction of the corresponding kinetic theory of spin-charge scattering processes detailed in Sec. \ref{sec:spin-charge}. To the large extend we follow here Giamarchi textbook Ref.~\cite{Giamarchi} for the bosonization procedure and notations, plus the Haldane review article Ref.~\cite{Haldane} to include band curvature effects. An additional element presented here is a more detailed bosonization of the interaction part of the fermionic Hamiltonian. It will be shown that similar to the band-curvature terms, interaction also generates anharmonic couplings between the spin and charge modes. As discussed in the main text of the paper these nuances have important consequences for the relaxation in 1D including spin-charge drag and energy transport.

The starting point is the usual form of fermionic Hamiltonian:
\begin{align}\label{eq:H}
&H\!=\!-i v_F\!\sum_{s}\!\!\int\!\! dx\!
\left[\psi^\dag_{Rs}(x)\partial_x\psi_{Rs}(x)-\psi^\dag_{Ls}(x)\partial_x\psi_{Ls}(x)\right]
\nonumber\\
&-\frac{1}{2m}\sum_{s}\!\int\! dx\!
\left[\psi^\dag_{Rs}(x)\partial^2_x\psi_{Rs}(x)+\psi^\dag_{Ls}(x)\partial^2_x\psi_{Ls}(x)\right]
\nonumber\\
&+\frac{1}{2}\sum_{ss'}\!\int\! dxdx'
V(x-x')\psi^\dag_s(x)\psi^\dag_{s'}(x')\psi_{s'}(x')\psi_s(x)\,.
\end{align}
Here index $s=\uparrow\downarrow$ stands for the spin projection, $\psi_{Rs}$ and $\psi_{Ls}$ are the annihilation operators for right- and left-moving spin-$s$ electrons, while $\psi_{s}=\psi_{Rs}+\psi_{Ls}$ is full operator in the interaction part of the Hamiltonian. The standard approximation is that low energy excitations take place near the Fermi points, such that electron operator is decomposed as follows
\begin{equation}
\psi_s(x)=\psi_{Rs}(x)+\psi_{Ls}(x)=e^{ik_Fx}R_s(x)+e^{-k_Fx}L_s(x)
\end{equation}
where new fields $R(L)_s(x)$ are assumed to vary slowly on the scale of the Fermi wavelength. In the bosonization description these
fields can be expressed in terms of dosonic displacement $\varphi_s(x)$ and conjugated phase $\vartheta_s(x)$
\begin{align}
R_s(x)=\frac{\kappa_s}{\sqrt{2\pi a}}\exp[i\vartheta_s(x)-i\varphi_s(x)]\, \nonumber\\ 
L_s(x)=\frac{\kappa_s}{\sqrt{2\pi a}}\exp[i\vartheta_s(x)+i\varphi_s(x)]\,,
\end{align}
where $a$ is the short distance cut-off $\sim k^{-1}_{F}$ and $\kappa_s$ are the Klein factors that ensure proper anticommutation relation between original fermionic operators. They obey $\{\kappa_s,\kappa_{s'}\}=2\delta_{ss'}$ and satisfy $\kappa^\dag_s=\kappa_s$. The bosonic fields obey commutation
\begin{equation}
[\varphi_s(x),\vartheta_{s'}(x)]=\frac{i\pi}{2}\sign(x-x')\delta_{ss'}
\end{equation}
With these notations at hand fermionic densities for right- and left-moving electrons become
\begin{align}
\rho_{Rs}(x)=R^\dag_{s}(x)R_{s}(x)=-\frac{1}{2\pi}\partial_x[\varphi_s(x)-\vartheta_s(x)],\nonumber \\
\rho_{Ls}(x)=L^\dag_{s}(x)L_{s}(x)=-\frac{1}{2\pi}\partial_x[\varphi_s(x)+\vartheta_s(x)].
\end{align}
The total density operator per spin $\rho_s(x)=\psi^\dag_s(x)\psi_s(x)$ contains the sum of the long-wavelenght part, $\rho^{(0)}_s(x)$, and oscillatory part $\rho^{(2k_F)}_s(x)$:
\begin{align}
&\rho_s(x)=\rho^{(0)}_s(x)+\rho^{(2k_F)}_s(x)=\nonumber \\ 
&-\frac{1}{\pi}\partial_x\varphi_s(x)+\frac{1}{\pi a}\cos[2\varphi_s(x)-2k_Fx]\,.
\end{align}
For the first two terms of Eq.~\eqref{eq:H} in the bosonization dictionary we have
\begin{align}
\psi^\dag_{Rs}(x)\partial_x\psi_{Rs}(x)-\psi^\dag_{Ls}(x)\partial_x\psi_{Ls}(x)=\nonumber \\ 
i\pi[\rho^2_{Rs}(x)+\rho^2_{Ls}(x)]\,,\nonumber \\
\psi^\dag_{Rs}(x)\partial^2_x\psi_{Rs}(x)+\psi^\dag_{Ls}(x)\partial^2_x\psi_{Ls}(x)= \nonumber\\
-\frac{4\pi^2}{3}[\rho^3_{Rs}(x)+\rho^3_{Ls}(x)]\,,
\end{align}
such that kinetic part of the Hamiltonian transforms into
\begin{align}
&H_{\text{kin}}=\frac{v_F}{2\pi}\sum_s\int dx
\left[(\partial_x\varphi_s)^2+(\partial_x\vartheta_s)^2\right]\nonumber \\ 
&-\frac{1}{6\pi m}\sum_s\int
dx\left[(\partial_x\varphi_s)^3+3(\partial_x\varphi_s)(\partial_x\vartheta_s)^2\right].
\end{align}

For the interaction part of the Hamiltonian in Eq. \eqref{eq:H} we proceed as follows. Up to an additive constant it can be rewritten as
\begin{align}
&H_{int}=\frac{1}{2}\sum_{ss'}\int dxdx' V(x-x') \nonumber \\
&[\underbrace{\psi^\dag_{Rs}(x)\psi_{Rs}(x)+
\psi^\dag_{Ls}(x)\psi_{Ls}(x)}_{q\sim0}+
\underbrace{\psi^\dag_{Ls}(x)\psi_{Rs}(x)}_{q\sim2k_F}]\nonumber\\
&[\underbrace{\psi^\dag_{Rs'}(x')\psi_{Rs'}(x')
+\psi^\dag_{Ls'}(x')\psi_{Ls'}(x')}_{q\sim0}+
\underbrace{\psi^\dag_{Ls'}(x')\psi_{Rs'}(x')}_{q\sim2k_F}]
\end{align}
thus separating explicitly different scattering channels, where $q$ labels characteristic momenta transferred in the collision. From
here one can read out forward and backward scattering parts of the interaction term, namely $H_{\text{int}}=H^{\text{fs}}_{\text{int}}+H^{\text{bs}}_{\text{int}}$. The formed one may be easily rewritten in the bosonization dictionary
\begin{align}
H^{\text{fs}}_{\text{int}} =\frac{1}{2}\sum_{ss'}\int dxdx'
V(x-x')\rho^{(0)}_{s}(x)\rho^{(0)}_{s'}(x')\nonumber \\ 
=\frac{V_0}{2\pi^2}\sum_{ss'}\int dx(\partial_x\varphi_s)(\partial_x\varphi_{s'}).
\end{align}
Here we expanded the density $\rho^{(0)}_{s'}(x')$ around $x$ using the fact that interaction potential $V(x-x')$ is short-ranged
and that fields $\varphi_{s}(x)$ are slowly varying on the scale where interaction appreciably decays. By $V_0$ we denote zero
momentum Fourier transform of the interaction potential and higher order gradients were ignored. We concentrate now on the backward
scattering part of the Hamiltonian. In the bosonization language it can be written as
\begin{align}
&H^{\text{bs}}_{\text{int}}=\frac{1}{8\pi^2a^2}\sum_{ss'}\int dxdy\,
V(y)\nonumber \\ 
&\times\left[e^{-2ik_Fy}e^{2i\varphi_s(x+y/2)-2i\varphi_{s'}(x-y/2)}+h.c.\right].
\end{align}
For the case of $s\neq s'$ we have sine-Gordon part of the Hamiltonian
\begin{equation}
H^{\text{bs}|ss'}_{\text{int}}=\frac{V_{2k_F}}{2\pi^2a^2}\int dx
\cos[2\varphi_\uparrow(x)-2\varphi_\downarrow(x)]
\end{equation}
A little more careful consideration is required for the case when $s=s'$. In this case one should expand the fields, which
are in fact non-commuting operators. To do the expansion procedure safely the operator has to be normal-ordered:
\begin{align}
\exp[i\varphi_s(x)]=\langle\exp[i\varphi_s(x)]\rangle:\exp[i\varphi_s(x)]:\nonumber \\ 
=\exp[-\langle\varphi^2_s(x)\rangle/2]:\exp[i\varphi_s(x)]:
\end{align}
such that one may apply usual Taylor series for the operator under the normal-ordered sign $:(\ldots):$. The brackets $\langle\ldots\rangle$ imply quantum averaging and to the lowest order in interaction
\begin{equation}
\langle(\varphi_s(x)-\varphi_s(x'))^2\rangle=\ln\left[\frac{|x-x'|}{a}\right]\,.
\end{equation}
With this formalism at hand we have
\begin{align}
H^{\text{bs}|ss}_{\text{int}}&\approx
\frac{1}{8\pi^2a^2}\sum_{s}\int dxdy
V(y)\left[e^{-2ik_Fy}e^{-2\ln\frac{|y|}{a}}\right. \nonumber \\ 
&\times\left(1+2i\Phi_s(x,y)+ \frac{(2i)^2}{2}\Phi^2_s(x,y)\right.\nonumber \\ 
&+\left.\left.\frac{(2i)^3}{6}\Phi^3_s(x,y)+
\frac{(2i)^4}{24}\Phi^4_s(x,y)\right)+h.c.\right]
\end{align}
where $\Phi_s(x,y)=\varphi_s(x+y/2)-\varphi_s(x-y/2)\approx y\partial_x\varphi_s(x)$ and we carried gradient expansion to
the lowest order. By neglecting now constant and full derivative terms and noticing that $\exp[-2\ln(|y|/a)]=(a/y)^2$, that
cancels cut-off dependent prefactor, one finds
\begin{align}
H^{\text{bs}|ss}_{\text{int}}\approx\frac{1}{8\pi^2}\sum_s\int dxdy V(y)
\left[-4\cos(2k_Fy)(\partial_x\varphi_s)^2\right. \nonumber \\ 
-\left.\frac{8}{3}y\sin(2k_Fy)(\partial_x\varphi_s)^3
+\frac{4}{3}y^2\cos(2k_Fy)(\partial_x\varphi_s)^4\right].
\end{align}
After the integration by parts above expression reduces to the form 
\begin{align}
&H^{\text{bs|}ss}_{\text{int}}=-\frac{V_{2k_F}}{2\pi^2}\sum_s\int
dx(\partial_x\varphi_s)^2\nonumber \\ 
&+\frac{V'_{2k_F}}{3\pi^2}\sum_s\int
dx(\partial_x\varphi_s)^3-\frac{V''_{2k_F}}{6\pi^2}\sum_s\int
dx(\partial_x\varphi_s)^4.
\end{align}
The first term is conventional for the bosonization technique while the last two are new additions responsible for the interaction of bosons. At this point we perform transformation to the spin-charge representation for the boson fields:
\begin{equation}
\varphi_\rho=\frac{1}{\sqrt{2}}(\varphi_\uparrow+\varphi_\downarrow)\,,\qquad
\varphi_\sigma=\frac{1}{\sqrt{2}}(\varphi_\uparrow-\varphi_\downarrow)\,,
\end{equation}
and similar for the $\vartheta$-field. The final result we split into five parts
\begin{equation}
H=H_2+H^{\text{bc}}_3+H^{\text{bs}}_{3}+H_4+H_{\text{sg}}. 
\end{equation}
The quadratic part
\begin{align}
H_2=\frac{v_\rho}{2\pi}\int
dx\left[\frac{1}{K_\rho}(\partial_x\varphi_\rho)^2+K_\rho(\partial_x\vartheta_\rho)^2\right]\nonumber \\ 
+\frac{v_\sigma}{2\pi}\int
dx\left[\frac{1}{K_\sigma}(\partial_x\varphi_\sigma)^2+K_\sigma(\partial_x\vartheta_\sigma)^2\right]
\end{align} 
corresponds to the usual linear Luttinger liquid model of spin-charge separation explicit. The boson velocities and Luttinger liquid interaction constants are locked by relations $v_\rho K_\rho=v_\sigma K_\sigma=v_F$, and at the level of perturbation theory $K_\rho=1-(2V_0-V_{2k_F})/2\pi v_F$ and $K_\sigma=1+V_{2k_F}/2\pi v_F$. The sine-Gordon term is also belongs to the linear Luttinger liquid theory 
\begin{equation}
H_{\text{sg}}=\frac{V_{2k_F}}{2\pi^2a^2}\int dx
\cos[2\sqrt{2}\varphi_\sigma],
\end{equation}
The cubic order coupling terms can be split into two groups. First group is due to band curvature  
\begin{align}
H^{\text{bc}}_{3}&=-\frac{1}{6\sqrt{2}\pi m}\int
dx\left[(\partial_x\varphi_\rho)^3\right. \nonumber \\ 
&+3(\partial_x\varphi_\rho)(\partial_x\varphi_\sigma)^2 +3(\partial_x\varphi_\rho)(\partial_x\vartheta_\rho)^2\nonumber \\ 
&\left.+3(\partial_x\varphi_\rho)(\partial_x\vartheta_\sigma)^2+6(\partial_x\vartheta_\rho)(\partial_x\varphi_\sigma)(\partial_x\vartheta_\sigma)\right]
\end{align}
The second group is due to backscattering  
\begin{equation}
H^{\text{bs}}_{3}=\frac{V'_{2k_F}}{3\sqrt{2}\pi^2}\int
dx\left[(\partial_x\varphi_\rho)^3+3(\partial_x\varphi_\rho)(\partial_x\varphi_\sigma)^2\right]. 
\end{equation}
These terms couple spin and charge excitations and lead to $\rho\to\sigma\sigma$ decay processes that we discussed in the context of spin-charge drag equilibration rates. The quartic order terms  
\begin{equation}
H_{4}\!=\!-\frac{V''_{2k_F}}{12\pi^2}\!\!\int\!\!
dx\left[(\partial_x\varphi_\rho)^4+6(\partial_x\varphi_\rho)^2(\partial_x\varphi_\sigma)^2
+(\partial_x\varphi_\sigma)^4\right]
\end{equation}
lead to $\rho\rho$, $\rho\sigma$, and $\sigma\sigma$  type boson scattering. Finally, to obtain kinetic equations for bosons 
we use canonical oscillator representation in normal modes 
\begin{align}
&\partial_x\varphi_{\rho/\sigma}=-\sum_q\sqrt{\frac{\pi|q|}{2L}}e^{-iqx}\left[b^\dag_{\rho/\sigma}(q)+b_{\rho/\sigma}(-q)\right], \nonumber \\ 
&\partial_x\vartheta_{\rho/\sigma}=\sum_q\sqrt{\frac{\pi|q|}{2L}}\sign(q)e^{-iqx}\left[b^\dag_{\rho/\sigma}(q)-b_{\rho/\sigma}(-q)\right],
\end{align}
written in terms of creation and annihilation operators. It should be also borne in mind that fields $\varphi_{\rho/\sigma}(x)$ and $\vartheta_{\rho/\sigma}(x)$ contain topological terms, $N^R\pm N^L$, which are important in defining the momentum operator \cite{Haldane}.  

\section{Mobile impurity model}

The purpose of this section is to illustrate a connection between calculation of quasiparticle decay rates in fermions via three-particle collisions and in bosons via nonlinear Luttinger liquid approach. For simplicity we condense this discussion to the spinless case. 
Here we essentially follow the framework developed in Ref. \cite{Imambekov-Glazman} with an extension to include an additional interaction term known from the context of impurity dynamics in Luttinger liquid \cite{CastroNeto-Fisher} that enables a decay processes. 

The technical essence of the method can be summarized as follows. Starting from the initial fermionic model one introduces not only conventional low-energy sub-bands $\psi_{R(L)}$ at $\pm k_F$ for right-movers and left-movers, but also the sub-band modes $d$ around the momentum $k$ of the high-energy particle (or hole) whose energy defines the threshold. In this approach, the fermion operator is split as $\psi(x)\sim e^{ik_Fx}\psi_R(x)+e^{-k_Fx}\psi_L(x)+e^{ikx}d(x)$ in which the high-energy particle acts as a mobile impurity coupled to the Luttiger liquid modes. The Hamiltonian for this model
reads~\cite{Imambekov-Glazman}
\begin{subequations}\label{eq:H-NLL}
\begin{align}
&H=H_0+H_d+H_{int}\,,\\
&H_0=\frac{v}{2\pi}\int
dx\left[K(\partial_x\vartheta)^2+K^{-1}(\partial_x\varphi)^2\right]\,,\\
&H_d=\int dx\,
d^\dag(x)\left[\varepsilon(k)-iv_d\partial_x\right]d(x)\,,\\
&H_{\text{int}}=\frac{1}{2\pi}\int dx \left[(V_R-V_L)\partial_x\vartheta-(V_R+V_L)\partial_x\varphi\right]\rho_d\,.
\end{align}
\end{subequations}
Here operator $d(x)$ creates a mobile particle of momentum $k$ and velocity $v_d = \partial\varepsilon/\partial k$,
and $\rho_d(x) = d^\dag(x)d(x)$ is the fermion density operator. In this treatment the curvature of the fermion dispersion was kept explicit.  
When applied to the calculation of the spectral function, this model captures power-law threshold singularities beyond the limit of linear Luttinger liquid theory and yields the universal description. However, this model does not yet capture relaxation processes as the interaction term couples mobile particle either to the left-movers or to the right-movers separately. At the level of fermionic description of the problem, we saw that finite decay rate is generated by RRL process that involves particle and two particle-hole pairs. This suggests that we need another coupling term of $d$-particle with both left- and right-movers. We thus add 
\begin{equation}
H'_{\text{int}}=\gamma\int dx\, \rho_d(\partial_x\varphi_R)(\partial_x\varphi_L)
\end{equation}    
which is inspired by Ref.~\cite{CastroNeto-Fisher} where friction of a heavy particle moving through the Luttinger liquid was considered. An estimate for the coupling constant was given $\gamma\sim V^2/\varepsilon_F$ which is qualitatively consistent with three-particle scattering process.

Let us return now to the effective hamiltonian Eq. \eqref{eq:H-NLL} and look for a single high-energy particle. From $H_d$ the time-ordered (retarded) free propagator of $d$-electron is $G^{\text{ret}}_0(x,t)=\langle Td^\dag(x,t)d(0,0)\rangle=\theta(t)e^{i\varepsilon t}\delta(x-v_dt)$ which simply describes ballistically propagating particle. The Fourier transform of the latter is $G^{\text{ret}}_0(k,\omega)=(\omega-\varepsilon-kv_d+i\alpha)^{-1}$. The idea now is to apply perturbation theory in $H'_{\text{int}}$ to determine the self-energy for the $d$-particle induced
by collisions with right-movers and left-movers. 

The Dyson equation for the $d$-electron gives dressed propagator
\begin{equation}
G^{\text{ret}}(k,\omega)=\frac{1}{\omega-\varepsilon-kv_d-\Sigma^{\text{ret}}(k,\omega)}
\end{equation}
where self-energy appears to the second order in $\gamma$
\begin{equation}
\Sigma^{\text{ret}}=-i\gamma^2\!\!\int\!\! dxdt e^{-ikx+i\omega t}
\Pi_R(x,t)\Pi_L(x,t)G^{\text{ret}}_0(x,t)
\end{equation}
where free propagator for bosonic fields is
\begin{align}
\Pi_{R(L)}(x,t)=\langle\partial_{x}\varphi_{R(L)}(x,t)\partial_{x}\varphi_{R(L)}(0,0)\rangle\nonumber \\ 
=-\frac{1}{2\pi(x\mp vt\pm i\alpha)^2}.
\end{align}
A slight comment here that following Ref.~\cite{Imambekov-Glazman} it is a good idea to rescale bosonic fields first $\varphi\to\sqrt{K}\varphi$ and $\vartheta\to\vartheta/\sqrt{K}$, which removes LL interaction parameter $K$ from $H_0$ but renormalizes coefficients in $H'_{\text{int}}$. We reabsorbed all factors in the redefinition of $\gamma$ and so $\Pi_{R(L)}(x,t)$ is written above for the rescaled fields.
The particle life-time is determined by the imaginary part of the self-energy, so that we look at the latter
\begin{equation}
\mathrm{Im}\Sigma^{\text{ret}}=-\frac{\gamma^2}{4\pi^2}\int\limits^{+\infty}_{-\infty}dt
\frac{e^{i(\omega-\varepsilon-kv_d)t}}{[(v_d-v)t+i\alpha]^2[(v_d+v)t-i\alpha]^2}
\end{equation}
Notice here that by assumption $v_d>v$ so that poles of the
integrand are in the different parts of the complex plane. At the
mass-shell we find $d$-particle relaxation rate
\begin{equation}
\tau^{-1}_{d}=\mathrm{Im}\Sigma^{\text{ret}}(k,\omega=\varepsilon+v_dk)=\frac{\gamma^2(v^2_d-v^2)}{8\pi
v^3_d\alpha^3}
\end{equation}
We arrive at the finite result which, however, depends on the cut-off parameter $\alpha$. To resolve this issue we appeal to the fact that for the model to be well-defined there has to be a clear energy separation between the sub-bands of $d$-particle and bosonized right- and left-movers. One can argue from the dimensional analysis that cut-off should scale at low energies, $k\to k_F$, as $\alpha^{-1}\sim k-k_F$
Furthermore, one can take $v^2_d-v^2\approx 2v_F(k-k_F)/m^*$, and also $\gamma \propto V_{2k_F}(V_0-V_{2k_F})/(m^* v^2_F)$, where parametrically the latter is known from the perturbation theory in fermions of Sec. \ref{sec:qp-decay}. Finally, combining everything together we get
\begin{equation}
\tau^{-1}_d\sim\varepsilon_F \frac{V^2_{2k_F}(V_0-V_{2k_F})^2}{
v^4_F}\left(\frac{k-k_F}{k_F}\right)^4. 
\end{equation}
This estimate matches with the zero-temperature quasiparticles relaxation rate from Eq. \eqref{eq:tau-spinless-C}. 


\end{document}